\documentclass[english]{article}
\usepackage{lmodern}

\usepackage[T1]{fontenc}
\usepackage[latin9]{inputenc}
\usepackage{color}
\usepackage{mathrsfs}
\usepackage{amsmath}
\usepackage{amssymb}
\usepackage{colortbl}
\usepackage{booktabs}
\usepackage{multirow} 

\makeatletter

\usepackage{mathrsfs}   
\usepackage{slashed}     
\usepackage{bbold}  
\usepackage{url}
\usepackage{graphicx}
\usepackage[colorlinks=true,linkcolor=redLinks,citecolor=greenLinks,urlcolor=redLinks, pdfborder={0 0 1}]{hyperref}
\usepackage{xcolor}
\usepackage{framed}
\usepackage[numbers,sort&compress]{natbib}
\usepackage{amsmath}

\allowdisplaybreaks

\definecolor{greenLinks}{rgb}{0, 0.6, 0} 
\definecolor{blueLinks}{rgb}{0, 0, 0.6}
\definecolor{redLinks}{rgb}{0.6, 0, 0}

\usepackage{multirow}
\newcommand{\ol}{16\pi^2\,}
\newcommand{\lag}{\mathscr{L}}

\newcommand{\afive}[2]{\big(a^{(5)}_{#1}\big)_{#2}}
\newcommand{\rfive}[2]{\big(r^{(5)}_{#1}\big)_{#2}}
\newcommand{\afivep}[2]{\big(a^{(5)~\prime}_{#1}\big)_{#2}}
\newcommand{\rfivep}[2]{\big(r^{(5)~\prime}_{#1}\big)_{#2}}
\newcommand{\afived}[2]{\big(\dot{a}^{(5)}_{#1}\big)_{#2}}
\newcommand{\asix}[2]{\big(a^{(6)}_{#1}\big)_{#2}}
\newcommand{\rsix}[2]{\big(r^{(6)}_{#1}\big)_{#2}}
\newcommand{\asixp}[2]{\big(a^{(6)~\prime}_{#1}\big)_{#2}}
\newcommand{\asixd}[2]{\big(\dot{a}^{(6)}_{#1}\big)_{#2}}
\newcommand{\rsixp}[2]{\big(r^{(6)~\prime}_{#1}\big)_{#2}}

\newcommand{\tr}[1]{\mathrm{Tr}[#1]}

\newcommand{\ferbil}[2]{\Big[ #1 \Big]_{#2}}

\definecolor{RFcolor}{rgb}{0.48,0,0.76}

\makeatother

\usepackage{babel}
\begin{document}
\title{Renormalization of general Effective Field Theories: 
\\Renormalization of fermionic operators}
\author{Renato M. Fonseca${}^a$, Pablo Olgoso${}^{b,c}$ and Jos\'e Santiago${}^a$}

\maketitle

\begin{center}
{\Large{}\vspace{-0.5cm}}
${}^a$Departamento de F\'isica Te\'orica y del Cosmos,
Universidad de Granada, \\
Campus de Fuentenueva, E--18071 Granada,
Spain
~\\
${}^b$ Dipartimento di Fisica e Astronomia, Universit\`a di Padova, Via F. Marzolo 8, 35131 Padova, Italy
~\\
${}^c$ Istituto Nazionale di Fisica Nucleare, Sezione di Padova, Via F. Marzolo 8, 35131 Padova, Italy
~\\
Emails: renatofonseca@ugr.es, pablo.olgosoruiz@unipd.it, jsantiago@ugr.es
\par\end{center}

\begin{abstract} 
Renormalization group equations play a central role in effective field theories, both maintaining perturbative control and allowing one to determine the correct low-energy phenomenology. In this work, we complete the one-loop renormalization of the recently developed general effective field theory up to mass dimension six by providing the beta functions for physical fermionic operators. Together with Ref.~\cite{Fonseca:2025zjb}, our results allow one to renormalize any effective theory up to dimension six at one loop using only a group-theoretical calculation.

\end{abstract}

\section{Introduction}
Effective Field Theories (EFTs) are ubiquitous in the study of physical systems with separate scales. In quantum field theory, the large logarithms stemming from the ratio of these scales in quantum corrections can spoil the use of perturbation theory unless they are re-summed by means of the renormalization group equations (RGEs). Moreover, taking into account the mixing of operators at different energy scales is crucial to correctly evaluate the phenomenology of a model.
At one-loop order, the beta functions for LEFT and SMEFT are already known \cite{Jenkins:2017dyc,Jenkins:2013zja,Jenkins:2013wua,Alonso:2013hga} and automated \cite{Fuentes-Martin:2020zaz,DiNoi:2022ejg}. For SMEFT, partial results are available at dimension eight \cite{Chala:2021pll,AccettulliHuber:2021uoa,DasBakshi:2022mwk,Helset:2022pde,Assi:2023zid, Boughezal:2024zqa, Bakshi:2024wzz, Liao:2024xel}.  
Beta functions are also known for the ALP-LEFT and ALP-SMEFT \cite{Chala:2020wvs,Bauer:2020jbp,Bonilla:2021ufe}.
At two-loop order, several results  are already available for LEFT \cite{Naterop:2024cfx, Naterop:2025lzc,Naterop:2025cwg} and SMEFT \cite{DiNoi:2024ajj, Born:2024mgz,DiNoi:2025tka, Haisch:2025lvd, Haisch:2025vqj,Chala:2025crd,Guedes:2025sax}.
Furthermore, the calculation for a specific EFT at one loop is fully automated, thanks to tools like \texttt{Matchmakereft}~\cite{Carmona:2021xtq} or \texttt{Matchete}~\cite{Fuentes-Martin:2022jrf}. 

Nevertheless, the calculation remains tedious and must be repeated for every EFT.
This problem, which hampers both processes of matching and running,
compromises the overall efficiency of the EFT strategy in searches for new physics.
One approach that mitigates this issue is the construction of infrared/ultraviolet (IR/UV) dictionaries \cite{delAguila:2000rc,delAguila:2008pw,delAguila:2010mx, deBlas:2014mba, deBlas:2017xtg, Guedes:2023azv, Guedes:2024vuf}, which streamline the connection between low-energy observables and their ultraviolet origins.

An alternative approach, following the seminal works in \cite{Machacek:1983tz,Machacek:1983fi,Machacek:1984zw, Martin:1993zk, Luo:2002ti}, is to perform the calculations in a completely general EFT, with no restriction on field content or symmetry group. In this way, obtaining the results for any specific EFT reduces to a simple group-theoretical calculation.
In a previous work~\cite{Fonseca:2025zjb}, 
we defined the most general EFT up to mass dimension 6 both off-shell (Green's basis) and on-shell (physical basis), together with the explicit reduction of the former to the latter, and
we computed the one-loop beta functions for all bosonic operators in 
this general EFT, including both (physical) bosonic and fermionic operators as boundary conditions.
Similar works following the same idea appeared shortly after \cite{Misiak:2025xzq,Aebischer:2025zxg}, reproducing some of our results. 

In this companion paper we compute
the beta functions for the fermionic operators, thus completing the full renormalization of a general 
EFT up to mass-dimension six. The paper is structured as follows. In Section \ref{sec_general_EFT}, we review the  
structure of the general effective Lagrangian up to mass-dimension six. In Section \ref{sec_beta_functions} we provide the central result of our work, the beta functions for fermionic operators. We conclude in Section \ref{sec_concl} and present, for completeness, the results for the off-shell, one-loop divergences in Appendix \ref{matching:green}.

\section{The general effective Lagrangian to mass-dimension 6 \label{sec_general_EFT}}

The most general EFT up to mass dimension 6 was defined, both on-shell (physical basis) and off-shell (Green's basis) in a recent work~\cite{Fonseca:2025zjb}, together with the one-loop renormalization of the bosonic operators in 
the physical basis (sourced by both bosonic and fermionic operators).
In this work we are interested in the renormalization of the fermionic operators but, for clarity, we reproduce here the complete basis. 

Let us introduce our notation for the fields and symmetries defining our general EFT. We will consider a general gauge group $G$, which can be any compact Lie group, with structure constants $f^{ABC}$ and gauge bosons $V_\mu^A$. The field strength tensor reads
\begin{equation}
    F^A_{\mu\nu}=\partial_\mu V^A_\nu - \partial_\nu V^A_\mu + g_A f^{ABC} V^B_\mu V^C_\nu,
\end{equation}
and its covariant derivative
\begin{equation}
D_\rho F^A_{\mu\nu}\equiv (D_\rho F_{\mu\nu})^A = \partial_\rho F^A_{\mu\nu} + g_A f^{ABC} V^B_\rho F^C_{\mu\nu}.    
\end{equation}
All fermions --- taken to be left-handed (LH) --- can be collected in a single vector $\psi_i$ which transforms under some representation  of $G$, which is in general reducible. Likewise, all scalars $\phi_i$ are considered real and grouped in a single (potentially reducible) representation of $G$. Our convention for the covariant derivatives applied to these fields reads
\begin{align}
    D_\mu \psi_i &= \partial_\mu \psi_i - \mathrm{i} g_A t^A_{ij} V^A_\mu \psi_j, \\
    D_\mu \phi_a &= \partial_\mu \phi_a - \mathrm{i} g_A \theta^A_{ab} V^A_\mu \phi_b.
\end{align}
Given that gauge transformations preserve the norm of fields, both  the $t^A$ and the $\theta^A$ matrices must be hermitian. Additionally, since we are working with real scalars,  the $\theta^A$ must be purely imaginary and therefore anti-symmetric.

The renormalizable Lagrangian reads
\begin{align}
    \lag_{d\leq 4} =& -\frac{1}{4} (a_{KF})_{AB} F^A_{\mu\nu} F^{B\, \mu\nu}+ \frac{1}{2} (a_{K\phi})_{ab} D_\mu \phi_a D^\mu \phi_b
    + (a_{K\psi})_{ij} \bar{\psi}_i \mathrm{i} \slashed{D} \psi_j
    -\frac{1}{2} \Big[ (m_\psi)_{ij} \psi^T_i C \psi_j  + \mathrm{h.c.}\Big] 
\nonumber \\
&    -\frac{1}{2} (m_\phi^2)_{ab} \phi_a \phi_b 
-\eta_a \phi_a -\frac{1}{2} \Big[ Y_{ija} \psi^T_i C \psi_j  + \mathrm{h.c.}\Big] \phi_a
    -\frac{\kappa_{abc}}{3!} \phi_a \phi_b \phi_c
    -\frac{\lambda_{abcd}}{4!} \phi_a \phi_b \phi_c \phi_d,\label{L1}
\end{align}
where $C$ is the charge conjugation matrix.

At mass-dimension 5 we have,
\begin{align}
    \lag_5^{\mathrm{phys}} &=
    \bigg[ \frac{1}{2} \afive{\psi F}{Aij} \psi_i^T C \sigma^{\mu \nu} \psi_j F^A_{\mu\nu} 
    +\frac{1}{4} \afive{\psi \phi^2}{ijab}  \psi_i^T C \psi_j \phi_a \phi_b  
    + \mathrm{h.c.} \bigg]
    \nonumber \\
    &+\frac{1}{2} \afive{\phi F}{ABa} F^{A\, \mu\nu} F^B_{\mu\nu} \phi_a
    +\frac{1}{2} \afive{\phi \widetilde{F}}{ABa} F^{A\,\mu\nu} 
    \widetilde{F}^B_{\mu\nu} \phi_a 
    + \frac{1}{5!}\afive{\phi}{abcde} \phi_a \phi_b \phi_c \phi_d \phi_e, \label{L5phys}
    \\
    \lag_5^{\mathrm{red}}&=
    \frac{1}{2} \rfive{\phi \Box}{abc} (D_\mu D^\mu \phi_a) \phi_b \phi_c
    +\bigg[ 
    \frac{1}{2} \rfive{\psi}{ij} (D_\mu \psi_i)^T C D^\mu \psi_j
    +\rfive{\psi \phi}{ija} \bar{\psi}_i \mathrm{i} \slashed{D} \psi_j \phi_a
    +\mathrm{h.c.}\bigg ],\label{L5red}
\end{align}
 where we have used the standard definition:
\begin{equation}
    \sigma^{\mu\nu}=\frac{\mathrm{i}}{2} [\gamma^\mu,\gamma^\nu].
\end{equation}

At mass-dimension 6 we have:
\begin{align}
\lag_6^{\mathrm{phys}}&=
\frac{1}{3!} \asix{3F}{ABC} (F^A)^{~\nu}_\mu (F^B)^{~\rho}_\nu (F^C)^{~\mu}_\rho
+\frac{1}{3!} \asix{3\widetilde{F}}{ABC} (F^A)^{~\nu}_\mu (F^B)^{~\rho}_\nu (\widetilde{F}^C)^{~\mu}_\rho
\nonumber\\
&+ \frac{1}{4} \asix{\phi F}{ABab} F^{A}_{\mu\nu} F^{B\, \mu\nu} \phi_a \phi_b
+ \frac{1}{4} \asix{\phi\widetilde{F}}{ABab} F^{A}_{\mu\nu} \widetilde{F}^{B\, \mu\nu} \phi_a \phi_b
\nonumber \\
&+ \frac{1}{3} \asix{\phi D}{abcd} \left[(D_{\mu}\phi_{a})(D^{\mu}\phi_{b})\phi_{c}\phi_{d}+\left(ab\leftrightarrow cd\right)-\frac{1}{2}\left(a\leftrightarrow c\right)-\frac{1}{2}\left(a\leftrightarrow d\right)-\frac{1}{2}\left(b\leftrightarrow c\right)-\frac{1}{2}\left(b\leftrightarrow d\right)\right]
\nonumber \\
&+ \frac{1}{6!} \asix{\phi}{abcdef} \phi_a \phi_b \phi_c \phi_d \phi_e \phi_f
+
\frac{1}{2}\asix{\phi \psi}{ijab} \bar{\psi}_i \gamma^\mu \psi_j
[\phi_a D_\mu \phi_b - \phi_b D_\mu \phi_a]
+
\frac{1}{4} \asix{\bar{\psi}\psi}{ijkl}
(\bar{\psi}_i \gamma^\mu \psi_j)
(\bar{\psi}_k \gamma_\mu \psi_l)
\nonumber \\&
+ \bigg[ \frac{1}{2} \asix{\psi F}{Aija} F^A_{\mu\nu} \psi_i^T C \sigma^{\mu\nu} \psi_j \phi_a
+\frac{1}{2!3!} \asix{\psi \phi}{ijabc} \psi_i^T C \psi_j \phi_a \phi_b \phi_c
+ \frac{1}{4!} \asix{\psi \psi}{ijkl} (\psi_i^T C \psi_j) (\psi_k^T C \psi_l) +\mathrm{h.c.}\bigg], \label{L6phys}
\end{align}
\begin{align}
\lag_6^{\textrm{red}} & =
\frac{1}{2!}\rsix{2F}{AB} (D_\mu F^{A\, \mu \nu}) (D^\rho F^B_{\rho \nu})
+\frac{1}{2!}\rsix{FD\phi}{Aab}\left(D_{\nu}F^{A,\mu\nu}\right) \left[\left(D_{\mu}\phi_{a}\right)\phi_{b}-\left(a\leftrightarrow b\right)\right]
\nonumber \\
 & 
+\frac{1}{2!}\rsix{D\phi}{ab}\left(D_{\mu}D^{\mu}\phi_{a}\right)(D_{\nu}D^{\nu}\phi_{b})
 +\frac{1}{3!}\rsix{\phi D}{abcd}\left(D_{\mu}D^{\mu}\phi_{a}\right)\phi_{b}\phi_{c}\phi_{d}
 \nonumber \\
  & 
 +\rsix{D F\psi}{Aij} D^\nu F_{\mu\nu}^{A}
 \overline{\psi_{i}}\gamma^{\mu} \psi_{j} 
  +\rsix{F\psi}{Aij}F_{\mu\nu}^{A}\overline{\psi_{i}}\gamma^{\mu}
\mathrm{i}  \overleftrightarrow{D}^\nu
 \psi_{j} 
 +\rsix{\widetilde{F}\psi}{Aij}\widetilde{F}_{\mu\nu}^{A}\overline{\psi_{i}}\gamma^{\mu}
\mathrm{i}  \overleftrightarrow{D}^\nu
 \psi_{j} 
 \nonumber \\
 &+\frac{1}{2!}
 \rsix{\phi \psi 1}{ijab}
\left(\bar{\psi}_{i}i \overleftrightarrow{\slashed{D}}\psi_{j}\right)\phi_{a}\phi_{b}
+ \frac{1}{2!}\rsix{\phi \psi 2}{ijab} \overline{\psi}_{i}\gamma^{\mu}\psi_{j} D_{\mu}\left(\phi_{a}\phi_{b}\right)
 +\rsix{\psi D}{ij}i\overline{\psi}_{i}
 \{D_{\mu}D^{\mu},\slashed{D} \}\psi_{j}
\nonumber \\
 & +\left\{ \frac{1}{2!}\rsix{\psi\phi D1}{ija}
 \psi_{i}^{T}C\psi_{j}\left(D_{\mu}D^{\mu}\phi_{a}\right)
 +\frac{1}{2!}\rsix{\psi\phi D2}{ija}\left(D^{\mu}\psi_{i}\right)^{T}C i \sigma^{\mu\nu}\left(D_{\nu}\psi_{j}\right)\phi_{a}\right.\nonumber \\
 & +\left.\frac{1}{2!}\rsix{\psi\phi D3}{ija}\left(D_{\mu}D^{\mu}\psi_{i}\right)^{T}C\psi_{j}\phi_{a}
 +\textrm{h.c.}\right\}.  \label{L6red}
\end{align}
Finally, the following operators are redundant only in $d=4$, and consequently induce evanescent effects:
\begin{align}
    \mathcal{L}^{(6)}_{\mathrm{ev}}&=\frac{1}{4}(r_{\psi\overline{\psi}}^{(6)})_{ijkl}(\psi_i^TC\psi_j) (\overline{\psi}_k C\overline{\psi}^T_l)+\frac{1}{4}(r_{\psi \overline{\psi}2}^{(6)})_{ijkl}(\psi_i^TC\sigma_{\mu\nu}\psi_j) (\overline{\psi}_k \sigma^{\mu\nu}C\overline{\psi}^T_l)\\
    &+\bigg(\frac{1}{4!}(r_{\psi\psi 2}^{(6)})_{ijkl}(\psi_i^TC\sigma_{\mu\nu}\psi_j) (\psi_k^T C \sigma^{\mu\nu}\psi_l) + \mathrm{h.c.}\bigg).
\end{align}
These effects, however, do not modify the one-loop renormalization so we will not take them into account.

Further details, including the symmetry properties of the different WCs can be found in~\cite{Fonseca:2025zjb}.
In the following, whenever possible we will keep the fermionic indices implicit preserving the correct order in the matrix multiplication. Also the dagger symbol will include the transposition of fermionic indices.

\section{Results \label{sec_beta_functions}}

\subsection{General formalism}

For completeness, in this section we will summarize the procedure that we follow to obtain the beta functions of the general EFT. For further details, we refer to Appendix A of Ref.~\cite{Fonseca:2025zjb}.

Our starting point will be the physical basis of operators defined in Section \ref{sec_general_EFT} as a boundary condition in the UV, that schematically reads:
\begin{equation}
    \lag^{\mathrm{UV}}= \sum_{i} a_i \mathcal{O}_i, \label{ldiv}
\end{equation}
where $a_i$ denotes the Wilson coefficient of the physical operator $\mathcal{O}_i$. 
This Lagrangian will induce UV divergences that we regularize using Dimensional Regularization, working in $d=4-2\epsilon$ dimensions. For the calculation of these divergences we use background field method ~\cite{Abbott:1980hw}, and since it is performed off-shell, they are parametrized by the full (including non-trivial kinetic terms and redundant operators) EFT Lagrangian. These divergent Wilson Coefficients for the kinetic terms and bosonic operators were reported in Section 4.2 and Appendix B of Ref.~\cite{Fonseca:2025zjb}, and in this work we provide the ones for fermionic operators in Appendix~\ref{matching:green}.
These results have been obtained using \texttt{Matchmakereft}~\cite{Carmona:2021xtq}, implementing a general model and several specific toy models (whose WCs were computed with the help of \texttt{GroupMath}~\cite{Fonseca:2020vke}).
The reduction to the physical basis, implemented by the shifts provided in Ref.~\cite{Fonseca:2025zjb}, involves the off-shell divergences of some bosonic WCs. Therefore, to reproduce our results, some of the expressions from Ref.~\cite{Fonseca:2025zjb} are needed. 

After canonical normalization and reduction to the physical basis, the divergent Lagrangian reads,
\begin{equation}
    \lag^{\mathrm{div}} = \sum_{j} a^\prime_j(a) ~\mathcal{O}_j,
\end{equation}
where we have denoted with a prime the WCs that parameterize the UV divergences (which are a function of the WCs in the UV Lagrangian, as explicitly shown). 
These divergences can be canceled by splitting the WCs into renormalized WCs and counterterms:
\begin{equation}
a_i^{\mathrm{bare}}= \mu^{(n_i-2)\epsilon} (a_i + \delta a_i),
\end{equation}
where the scale $\mu$ is introduced to ensure that the renormalized WCs in $d$ dimensions have the same mass dimension as the bare ones in $d=4$. This fixes $n_i$ to the number of fields appearing in the operator $\mathcal{O}_i$.  
Since we have canonically normalized the divergent Lagrangian in Eq.\eqref{ldiv}, we then have:
\begin{equation}
\delta a_i = - a_i^\prime. \label{deltCieqCiprime}
\end{equation}

The corresponding counterterms fix the renormalization scale dependence of the renormalized WCs. At one loop they read:
\begin{equation}
\beta_i \equiv \mu \frac{\mathrm{d}a_i}{\mathrm{d}\mu} = -2 a_i^\prime. \label{beta_oneloop}
\end{equation}
We will report below the results in the form:
\begin{equation}
    \dot{a}_i \equiv 16 \pi^2 \beta_i.
\end{equation}

\subsection{Beta functions of the fermionic operators}
\begin{figure}[t]
    \centering
    \includegraphics[width=0.6\textwidth]{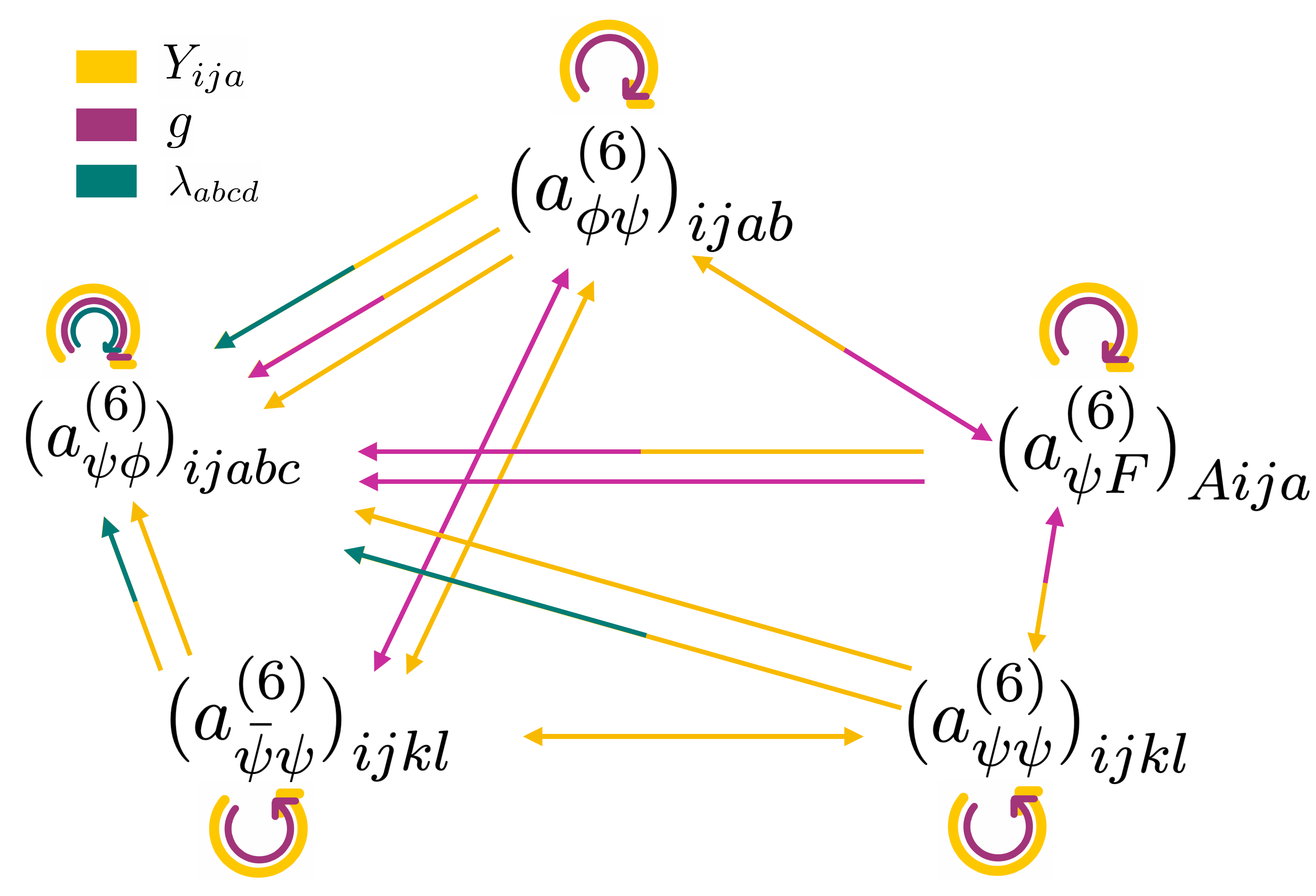}
    \caption{
    Renormalization structure of fermionic dimension-six operators, distinguishing contributions proportional to Yukawa $Y_{ija}$ (yellow), gauge $g$ (purple) or scalar quartic $\lambda$ (green) couplings. Double-colored arrows indicate contributions proportional to both couplings simultaneously.}
    \label{fig:ren_fermions}
\end{figure}

In this section we report the beta functions for the physical fermionic operators, which constitutes the main result of the article. Symmetrization of indices is either written explicitly ($\dot{a}_{ij}=c_{ij}+(i\leftrightarrow j)$ means $\dot{a}_{ij}=c_{ij}+c_{ji}$) or as a sum over permutations of all free indices, according to the symmetry properties of each coefficient as stated in Section \ref{sec_general_EFT}. As a visual guide, we depicted in Figure \ref{fig:ren_fermions} the structure of the mixing of the dimension-six fermionic operators under renormalization.

\begin{align}
(\dot{m}_\psi)_{ij} =&-3  
g^2\Big[m_\psi t^{A} t^{A}\Big]_{ij}  
+\frac{1}{2} \Big[m_\psi Y^\dagger_{a}  
Y_{a}\Big]_{ij} +\Big[Y_{a} m_\psi^\dagger  
Y_{a}\Big]_{ij} \nonumber\\
&-15 g\Big[m_\psi  
m_\psi^\dagger \afive{\psi F}{A}  
t^{A}\Big]_{ij} +3  
g\Big[m_\psi \afive{\psi F}{A}^\dagger  
m_\psi t^{A}\Big]_{ij} -\frac{1}{2}  
\afive{\psi \phi^2 }{ijab}(m_\phi^2)_{ab}  \nonumber\\
&
+\Big[\afive{\phi F}{ABa}\afive{\phi  
F}{ABb}+\afive{\phi  
\widetilde{F}}{ABa}\afive{\phi  
\widetilde{F}}{ABb}\Big]Y_{ija}\eta_{b}
\nonumber\\&
+4 g\Big[\afive{\phi  
F}{ABa}-\mathrm{i}\afive{\phi  
\widetilde{F}}{ABa}\Big]\eta_{a}\Big[\afive{\psi F}{A}  
t^{B}\Big]_{ij}  
\nonumber\\&
+6  
\eta_{a}\Big[\afive{\psi F}{A}  
Y^\dagger_{a} \afive{\psi F}{A}\Big]_{ij} -12  
\Big[\afive{\psi F}{A} m_\psi^\dagger  
m_\psi m_\psi^\dagger \afive{\psi  
F}{A}\Big]_{ij} \nonumber\\&-4  
\Big[m_\psi m_\psi^\dagger  
\afive{\psi F}{A} \afive{\psi  
F}{A}^\dagger m_\psi\Big]_{ij} -12  
\Big[m_\psi \afive{\psi F}{A}^\dagger  
m_\psi m_\psi^\dagger \afive{\psi  
F}{A}\Big]_{ij} \nonumber\\
&-  
\asix{\bar{\psi}  
\psi}{kilj}Y_{kla}\eta_{a} +2  
\asix{\bar{\psi}  
\psi}{kilj}\Big[m_\psi m_\psi^\dagger  
m_\psi\Big]_{kl} +\frac{1}{2}  
\asix{\psi  
\psi}{kilj}Y^\dagger_{kla}\eta_{a}  
\nonumber\\&+\frac{1}{2}  
\asix{\psi  
\psi}{ijkl}\Big[m_\psi^\dagger m_\psi  
m_\psi^\dagger\Big]_{kl}  + (i\leftrightarrow j) \\
\dot{Y}_{ija} =& 
+\frac{1}{4}   
\mathrm{Tr}\Big[Y_{a}   
Y^\dagger_{b}+Y_{b}   
Y^\dagger_{a}\Big]Y_{ijb}   
+\frac{1}{2}   
\Big[Y_{a} Y^\dagger_{b} Y_{b}\Big]_{ij}   
+\Big[Y_{b} Y^\dagger_{a} Y_{b}\Big]_{ij}   
-3 g^2\Big[Y_{a} t^{A}   
t^{A}\Big]_{ij}\nonumber\\&-6 g^2\Big[\afive{\phi   
F}{ABa}+\mathrm{i}\afive{\phi   
\widetilde{F}}{ABa}\Big]\Big[m_\psi t^{A} t^{B}\Big]_{ij} -18   
g\Big[m_\psi t^{A} Y^\dagger_{a}   
\afive{\psi F}{A}\Big]_{ij}   
\nonumber\\&
+6 g\Big[m_\psi   
Y^\dagger_{a} \afive{\psi F}{A}   
t^{A}\Big]_{ij} -12 g\Big[\afive{\psi F}{A} 
m_\psi^\dagger Y_{a} t^{A}\Big]_{ij} +3   
g\Big[\afive{\psi F}{A} m_\psi^\dagger   
(t^T)^{A} Y_{a}\Big]_{ij} \nonumber\\&-3   
g\theta^{A}_{ab}\Big[m_\psi \afive{\psi   
F}{A}^\dagger Y_{b}\Big]_{ij} +3 g\Big[m_\psi   
\afive{\psi F}{A}^\dagger Y_{a}   
t^{A}\Big]_{ij} -\frac{1}{2} \afive{\psi   
\phi^2 }{ijbc}\kappa_{abc}   
\nonumber\\&- \Big[m_\psi   
Y^\dagger_{b} \afive{\psi \phi^2   
}{ab}\Big]_{ij} -2 \Big[\afive{\psi   
\phi^2 }{ab} m_\psi^\dagger Y_{b}\Big]_{ij}  \nonumber\\&
+\Big[\afive{\phi F}{ABb}\afive{\phi   
F}{ABc}  
+\afive{\phi   
\widetilde{F}}{ABb}\afive{\phi   
\widetilde{F}}{ABc}\Big](m_\phi^2)_{ab}Y_{ijc} 
\nonumber\\&
+4   
g\Big[\afive{\phi   
F}{ABb}-\mathrm{i}\afive{\phi   
\widetilde{F}}{ABb}\Big](m_\phi^2)_{ab}\Big[\afive{\psi F}{A}   
t^{B}\Big]_{ij}   
\nonumber\\&
-8 g\Big[\afive{\phi   
F}{ABa} + 2 \mathrm{i}\afive{\phi   
\widetilde{F}}{ABa} \Big]\Big[m_\psi m_\psi^\dagger   
\afive{\psi F}{A} t^{B}\Big]_{ij} 
\nonumber\\&
-16   
g\Big[\afive{\phi F}{ABa}+\mathrm{i}\afive{\phi   
\widetilde{F}}{ABa}\Big]\Big[m_\psi   
\afive{\psi F}{A}^\dagger m_\psi   
t^{B}\Big]_{ij} \nonumber\\&
+6   
(m_\phi^2)_{ab}\Big[\afive{\psi F}{A}   
Y^\dagger_{b} \afive{\psi F}{A}\Big]_{ij} -12   
\Big[\afive{\psi F}{A} m_\psi^\dagger   
Y_{a} m_\psi^\dagger \afive{\psi   
F}{A}\Big]_{ij} \nonumber\\&-24   
\Big[\afive{\psi F}{A} Y^\dagger_{a}   
m_\psi m_\psi^\dagger \afive{\psi   
F}{A}\Big]_{ij} -2 \Big[m_\psi   
m_\psi^\dagger \afive{\psi F}{A}   
\afive{\psi F}{A}^\dagger Y_{a}\Big]_{ij} +4   
\Big[m_\psi m_\psi^\dagger Y_{a}   
\afive{\psi F}{A}^\dagger \afive{\psi   
F}{A}\Big]_{ij} \nonumber\\&-12   
\Big[m_\psi \afive{\psi F}{A}^\dagger   
Y_{a} m_\psi^\dagger \afive{\psi   
F}{A}\Big]_{ij} -2 \Big[m_\psi   
\afive{\psi F}{A}^\dagger \afive{\psi   
F}{A} m_\psi^\dagger Y_{a}\Big]_{ij} -12   
\Big[\afive{\psi F}{A} m_\psi^\dagger   
m_\psi \afive{\psi F}{A}^\dagger   
Y_{a}\Big]_{ij} \nonumber\\&-12   
\Big[m_\psi \afive{\psi F}{A}^\dagger   
m_\psi Y^\dagger_{a} \afive{\psi   
F}{A}\Big]_{ij} - \asix{\phi   
D}{abcd}(m_\phi^2)_{cd}Y_{ijb} +2   
\mathrm{i}(m_\phi^2)_{bc}\Big[Y_{c}   
\asix{\phi \psi}{ab}\Big]_{ij}   
\nonumber\\&-2   
\mathrm{i}\Big[m_\psi m_\psi^\dagger   
Y_{b} \asix{\phi \psi}{ab}\Big]_{ij} -2   
\mathrm{i}\Big[m_\psi \asix{\phi   
\psi}{ab} m_\psi^\dagger Y_{b}\Big]_{ij} +2   
\mathrm{i}\Big[Y_{b} m_\psi^\dagger   
m_\psi \asix{\phi \psi}{ab}\Big]_{ij}   
\nonumber\\&-   
\mathrm{i}\Big[m_\psi \asix{\phi   
\psi}{ab} Y^\dagger_{b} m_\psi\Big]_{ij} -3   
\mathrm{i}\Big[m_\psi Y^\dagger_{b}   
m_\psi \asix{\phi \psi}{ab}\Big]_{ij}   
- \asix{\bar{\psi}   
\psi}{kilj}(m_\phi^2)_{ab}Y_{klb}   
\nonumber\\&-2 \asix{\bar{\psi}   
\psi}{kjml}(m_\psi)_{ik}\Big[m_\psi^  
\dagger Y_{a}\Big]_{lm} +4   
\asix{\bar{\psi}   
\psi}{kilj}\Big[m_\psi m_\psi^\dagger   
Y_{a}\Big]_{kl} +2 \asix{\bar{\psi}   
\psi}{kjml}(m_\psi)_{ik}\Big[Y^\dagger_{a}  
m_\psi\Big]_{lm} \nonumber\\&+2   
\asix{\bar{\psi}   
\psi}{kilj}\Big[m_\psi Y^\dagger_{a}   
m_\psi\Big]_{lk} +12 g\Big[m_\psi   
m_\psi^\dagger (t^T)^{A} \asix{\psi   
F}{Aa}\Big]_{ij} -\frac{1}{2} \asix{\psi   
\phi}{ijabc}(m_\phi^2)_{bc}   
\nonumber\\&+\frac{1}{2}   
\asix{\psi   
\psi}{kilj}(m_\phi^2)_{ab}Y^\dagger_{klb}   
+\frac{1}{2} \asix{\psi   
\psi}{ijkl}\Big[m_\psi^\dagger Y_{a}   
m_\psi^\dagger
+2 m_\psi^\dagger  m_\psi   Y^\dagger_{a}
\Big]_{kl}   
+(i\leftrightarrow j),\\
\afived{\psi F}{Aij} =& +\Big[\frac{1}{6}   
f_{ACD}f_{BCD} +\frac{1}{12}   
\theta^{A}_{ab}\theta^{B}_{ba}   
+\frac{1}{3} \mathrm{Tr}\Big[t^{B}   
t^{A}\Big] \Big]g^2\afive{\psi F}{Bij}
\nonumber\\&
-7   
g^2\mathrm{i}f_{BAC}\Big[\afive{\psi F}{B}  
t^{C}\Big]_{ij} +g^2\Big[\afive{\psi F}{A}  
t^{B} t^{B}\Big]_{ij} \nonumber\\&+3   
g\Big[\afive{\psi F}{B} t^{B} t^{A}\Big]_{ij}   
+g^2\Big[\afive{\psi F}{B} t^{A}   
t^{B}\Big]_{ij} +\frac{1}{2}   
\Big[\afive{\psi F}{A} Y^\dagger_{a}   
Y_{a}\Big]_{ij}   
\nonumber\\&
+\Big[\afive{\phi   
F}{ABa}+\mathrm{i}\afive{\phi   
\widetilde{F}}{ABa}\Big]
\Big[m_\psi Y^\dagger_{a}   
\afive{\psi F}{B}\Big]_{ij}   
+2   
\Big[\afive{\phi F}{ABa}+\mathrm{i}\afive{\phi   
\widetilde{F}}{ABa}\Big]
\Big[\afive{\psi   
F}{B} m_\psi^\dagger Y_{a}\Big]_{ij}   
\nonumber\\&   
+\Big[\afive{\phi F}{ABa}+\mathrm{i}\afive{\phi   
\widetilde{F}}{ABa}]\Big[m_\psi   
\afive{\psi F}{B}^\dagger Y_{a}\Big]_{ij}   
\nonumber\\&+ g\afive{\psi   
F}{Bij}\mathrm{Tr}\Big[m_\psi^\dagger   
\afive{\psi F}{B} t^{A}+m_\psi^\dagger   
\afive{\psi F}{A} t^{B}\Big] -3   
g\Big[\afive{\psi F}{A} m_\psi^\dagger   
\afive{\psi F}{B} t^{B}\Big]_{ij}   
\nonumber\\&+g\afive{\psi   
F}{Bij}\mathrm{Tr}\Big[t^{A}   
\afive{\psi F}{B}^\dagger m_\psi + t^{B}   
\afive{\psi F}{A}^\dagger m_\psi\Big]   
+7   
g\Big[m_\psi \afive{\psi F}{B}^\dagger   
\afive{\psi F}{A} t^{B}\Big]_{ij}   
\nonumber\\&+2 g\Big[m_\psi   
\afive{\psi F}{B}^\dagger \afive{\psi   
F}{B} t^{A}\Big]_{ij} -3   
g\mathrm{i}f_{ACB}\Big[m_\psi   
\afive{\psi F}{C}^\dagger \afive{\psi   
F}{B}\Big]_{ij} \nonumber\\
&-\frac{g^2}{2}   
\Big[\mathrm{i}\asix{3F}{ABC}-\asix{3\widetilde{F}}{ABC}\Big]\Big[m_\psi t^{B}   
t^{C}\Big]_{ij}    
\nonumber\\&+\Big[m_\psi   
Y^\dagger_{a} \asix{\psi F}{Aa}\Big]_{ij} +2   
\Big[\asix{\psi F}{Aa} m_\psi^\dagger   
Y_{a}\Big]_{ij}   
\nonumber\\&-\frac{g}{6}   
\asix{\psi   
\psi}{ikjl}\Big[m_\psi^\dagger   
(t^T)^{A}\Big]_{lk} 
-(i\leftrightarrow j), \\
\afived{\psi \phi^2}{ijab} =&\sum_{\mathrm{perm}}\Big\{
+3 g^2\afive{\phi  
F}{ABb}\theta^{A}_{cd}\theta^{B}_{da}Y_{ijc} +6  
g^2\mathrm{i}\afive{\phi  
\widetilde{F}}{ABb}\Big[Y_{a} t^{B} t^{A}\Big]_{ij}  
-6 g^2\afive{\phi F}{ABb}\Big[(t^T)^{A} Y_{a}  
t^{B}\Big]_{ij} \nonumber\\&\phantom{\sum_{\mathrm{perm}}\Big\{}
+12  
g^3\theta^{A}_{ac}\theta^{B}_{cb}\Big[\afive{ 
\psi F}{A} t^{B}\Big]_{ij} +12  
g\Big[\afive{\psi F}{A} Y^\dagger_{a} Y_{b}  
t^{A}\Big]_{ij} -6 g\Big[\afive{\psi F}{A}  
t^{A} Y^\dagger_{a} Y_{b}\Big]_{ij}  
\nonumber\\&\phantom{\sum_{\mathrm{perm}}\Big\{}
-6 g\Big[\afive{\psi  
F}{A} Y^\dagger_{a} (t^T)^{A} Y_{b}\Big]_{ij}   
+\frac{1}{2} g^2\afive{\psi \phi^2  
}{ijcd}\theta^{A}_{ac}\theta^{A}_{bd}  
\nonumber\\&\phantom{\sum_{\mathrm{perm}}\Big\{}
- g^2\afive{\psi  
\phi^2 }{ijac}\theta^{A}_{bd}\theta^{A}_{dc}  
+\frac{1}{4} \afive{\psi \phi^2  
}{ijbc}\mathrm{Tr}\Big[Y_{a}  
Y^\dagger_{c} + Y_c Y_a^\dagger\Big] 
\nonumber\\&\phantom{\sum_{\mathrm{perm}}\Big\{}
+\frac{1}{4}  
\mathrm{Tr}\Big[\afive{\psi \phi^2  
}{ab} Y^\dagger_{c}\Big]Y_{ijc} +\frac{1}{4}  
\Big[\afive{\psi \phi^2 }{ab}  
Y^\dagger_{c} Y_{c}\Big]_{ij}  
+\Big[\afive{\psi \phi^2 }{ac}  
Y^\dagger_{c} Y_{b}\Big]_{ij}  
\nonumber\\&\phantom{\sum_{\mathrm{perm}}\Big\{}
+2 \Big[\afive{\psi  
\phi^2 }{ac} Y^\dagger_{b} Y_{c}\Big]_{ij}  
+\frac{1}{4} \afive{\psi \phi^2  
}{ijcd}\lambda_{abcd} -2  
g^2\theta^{A}_{bc}\Big[\afive{\psi  
\phi^2 }{ac} t^{A}\Big]_{ij}  
\nonumber\\&\phantom{\sum_{\mathrm{perm}}\Big\{}
+\frac{1}{2}  
g^2\Big[\afive{\psi \phi^2 }{ab} t^{A}  
t^{A}\Big]_{ij} +2 g^2\Big[(t^{A})^T \afive{\psi  
\phi^2 }{ab} t^{A}\Big]_{ij} +\frac{1}{4}  
\mathrm{Tr}\Big[\afive{\psi \phi^2  
}{ab}^\dagger Y_{c}\Big]Y_{ijc}  
\nonumber\\&\phantom{\sum_{\mathrm{perm}}\Big\{}
+\frac{1}{2} \Big[Y_{c}  
\afive{\psi \phi^2 }{ab}^\dagger  
Y_{c}\Big]_{ij} 
-\frac{1}{2}\Big[ 
\afive{\phi  
F}{ABc}\afive{\phi F}{ABd}
+\afive{\phi  
\widetilde{F}}{ABc}\afive{\phi  
\widetilde{F}}{ABd}
\Big]
Y_{ijc}\kappa_{abd}  
\nonumber\\&\phantom{\sum_{\mathrm{perm}}\Big\{}
+12 g^2\Big[
\afive{\phi  
F}{ABa}\afive{\phi F}{ACb}
+\mathrm{i}\afive{\phi  
F}{ABa}\afive{\phi  
\widetilde{F}}{ACb}
\Big]\Big[m_\psi t^{B} 
t^{C}\Big]_{ij} 
\nonumber\\&\phantom{\sum_{\mathrm{perm}}\Big\{}
+2 g\Big[
\afive{\phi  
F}{ABc} -\mathrm{i}\afive{\phi  
\widetilde{F}}{ABc}
\Big]\kappa_{abc}\Big[\afive{\psi F}{A}  
t^{B}\Big]_{ij} 
\nonumber\\&\phantom{\sum_{\mathrm{perm}}\Big\{}
-6 g\afive{\phi  
F}{ABa}\theta^{B}_{bc}\Big[m_\psi  
Y^\dagger_{c} \afive{\psi F}{A}\Big]_{ij} +24  
g\afive{\phi F}{ABb}\Big[m_\psi t^{B}  
Y^\dagger_{a} \afive{\psi F}{A}\Big]_{ij}  
\nonumber\\&\phantom{\sum_{\mathrm{perm}}\Big\{}
-8 g\afive{\phi  
F}{ABb}\Big[m_\psi Y^\dagger_{a}  
\afive{\psi F}{A} t^{B}\Big]_{ij} +6  
g\mathrm{i}\afive{\phi  
\widetilde{F}}{ABa}\Big[m_\psi t^{B}  
Y^\dagger_{b} \afive{\psi F}{A}\Big]_{ij} 
\nonumber\\&\phantom{\sum_{\mathrm{perm}}\Big\{}
+2  
g\mathrm{i}\afive{\phi  
\widetilde{F}}{ABb}\Big[m_\psi Y^\dagger_{a}  
\afive{\psi F}{A} t^{B}\Big]_{ij}  
+24 g\afive{\phi  
F}{ABb}\Big[\afive{\psi F}{A}  
m_\psi^\dagger Y_{a} t^{B}\Big]_{ij} 
\nonumber\\&\phantom{\sum_{\mathrm{perm}}\Big\{}
+16  
g\afive{\phi F}{ABb}\Big[\afive{\psi  
F}{A} m_\psi^\dagger (t^T)^{B} Y_{a}\Big]_{ij} -16  
g\mathrm{i}\afive{\phi  
\widetilde{F}}{ABa}\Big[\afive{\psi F}{A}  
m_\psi^\dagger (t^T)^{B} Y_{b}\Big]_{ij}  
\nonumber\\&
\phantom{\sum_{\mathrm{perm}}\Big\{}
-18 g\afive{\phi  
F}{ABa}\theta^{B}_{bc}\Big[m_\psi  
\afive{\psi F}{A}^\dagger Y_{c}\Big]_{ij} +24  
g\afive{\phi F}{ABa}\Big[m_\psi  
\afive{\psi F}{A}^\dagger Y_{b}  
t^{B}\Big]_{ij} 
\nonumber\\&\phantom{\sum_{\mathrm{perm}}\Big\{}
+6 g\mathrm{i}\afive{\phi  
\widetilde{F}}{ABa}\Big[m_\psi  
\afive{\psi F}{A}^\dagger Y_{b}  
t^{B}\Big]_{ij} 
-18  
g\mathrm{i}\afive{\phi  
\widetilde{F}}{ABa}\Big[m_\psi  
\afive{\psi F}{A}^\dagger (t^T)^{B}  
Y_{b}\Big]_{ij} 
\nonumber\\&\phantom{\sum_{\mathrm{perm}}\Big\{}
+12  
g^2\theta^{A}_{ac}\theta^{B}_{cb}\Big[\afive{ 
\psi F}{A} m_\psi^\dagger \afive{\psi  
F}{B}\Big]_{ij} +3  
\kappa_{abc}\Big[\afive{\psi F}{A}  
Y^\dagger_{c} \afive{\psi F}{A}\Big]_{ij}  
\nonumber\\&\phantom{\sum_{\mathrm{perm}}\Big\{}
+12 \Big[\afive{\psi  
F}{A} Y^\dagger_{a} m_\psi Y^\dagger_{b}  
\afive{\psi F}{A}\Big]_{ij} +24  
\Big[\afive{\psi F}{A} Y^\dagger_{a} Y_{b}  
m_\psi^\dagger \afive{\psi  
F}{A}\Big]_{ij} 
\nonumber\\&\phantom{\sum_{\mathrm{perm}}\Big\{}
-6  
g^2\theta^{A}_{bc}\theta^{B}_{ac}\Big[m_\psi
\afive{\psi F}{B}^\dagger \afive{\psi  
F}{A}\Big]_{ij} 
+12  
\Big[m_\psi \afive{\psi F}{A}^\dagger  
Y_{a} Y^\dagger_{b} \afive{\psi  
F}{A}\Big]_{ij} 
\nonumber\\&\phantom{\sum_{\mathrm{perm}}\Big\{}
-2 \Big[m_\psi  
\afive{\psi F}{A}^\dagger \afive{\psi  
F}{A} Y^\dagger_{a} Y_{b}\Big]_{ij} -2  
\Big[m_\psi Y^\dagger_{a} Y_{b}  
\afive{\psi F}{A}^\dagger \afive{\psi  
F}{A}\Big]_{ij} \nonumber\\&\phantom{\sum_{\mathrm{perm}}\Big\{}
+12  
\Big[\afive{\psi F}{A} Y^\dagger_{a}  
m_\psi \afive{\psi F}{A}^\dagger  
Y_{b}\Big]_{ij} +12 \Big[\afive{\psi F}{A}  
m_\psi^\dagger Y_{a} \afive{\psi  
F}{A}^\dagger Y_{b}\Big]_{ij} 
\nonumber\\&\phantom{\sum_{\mathrm{perm}}\Big\{}
-4  
\Big[\afive{\psi F}{A} \afive{\psi  
F}{A}^\dagger Y_{b} m_\psi^\dagger  
Y_{a}\Big]_{ij} -6  
g\Big[\afive{\psi F}{A} m_\psi^\dagger  
\afive{\psi \phi^2 }{ab} t^{A}\Big]_{ij}  
\nonumber\\&\phantom{\sum_{\mathrm{perm}}\Big\{}
+\frac{3}{2} g\Big[\afive{\psi F}{A}  
m_\psi^\dagger (t^T)^{A} \afive{\psi  
\phi^2 }{ab}\Big]_{ij} +3  
g\theta^{A}_{bc}\Big[m_\psi \afive{\psi  
\phi^2 }{ac}^\dagger \afive{\psi  
F}{A}\Big]_{ij} \nonumber\\&\phantom{\sum_{\mathrm{perm}}\Big\{}
-6  
g\Big[m_\psi t^{A} \afive{\psi \phi^2  
}{ab}^\dagger \afive{\psi F}{A}\Big]_{ij} +3  
g\theta^{A}_{ac}\Big[m_\psi \afive{\psi  
F}{A}^\dagger \afive{\psi \phi^2  
}{bc}\Big]_{ij} 
\nonumber\\&\phantom{\sum_{\mathrm{perm}}\Big\{}
-\frac{15}{2} g\Big[m_\psi  
\afive{\psi F}{A}^\dagger (t^T)^{A}  
\afive{\psi \phi^2 }{ab}\Big]_{ij}  
-3 g\Big[m_\psi  
\afive{\psi F}{A}^\dagger \afive{\psi  
\phi^2 }{ab} t^{A}\Big]_{ij} 
\nonumber\\&\phantom{\sum_{\mathrm{perm}}\Big\{}
-  
\Big[\afive{\psi \phi^2 }{ac}  
m_\psi^\dagger \afive{\psi \phi^2  
}{bc}\Big]_{ij} -\frac{1}{2} \Big[m_\psi  
\afive{\psi \phi^2 }{ac}^\dagger  
\afive{\psi \phi^2 }{bc}\Big]_{ij}  
\nonumber\\&\phantom{\sum_{\mathrm{perm}}\Big\{}
+\asix{\phi  
D}{cbde}Y_{ijc}\kappa_{ade} -\frac{3}{2}  
\asix{\phi D}{abcd}Y_{ije}\kappa_{ecd} -2  
\asix{\phi D}{acbd}\Big[m_\psi  
Y^\dagger_{d} Y_{c}\Big]_{ij}  
\nonumber\\&\phantom{\sum_{\mathrm{perm}}\Big\{}
+2 \asix{\phi  
D}{abcd}\Big[Y_{c} m_\psi^\dagger  
Y_{d}\Big]_{ij} 
+3 g^2\Big[
\asix{\phi  
F}{ABab} +\mathrm{i}\asix{\phi  
\widetilde{F}}{ABab}
\Big]\Big[m_\psi t^{A} t^{B}\Big]_{ij} 
\nonumber\\&\phantom{\sum_{\mathrm{perm}}\Big\{}
+\mathrm{i}\mathrm{Tr} 
\Big[\asix{\phi \psi}{ac}  
m_\psi^\dagger Y_{b}\Big]Y_{ijc} +2  
\mathrm{i}\Big[Y_{a} m_\psi^\dagger Y_{c}  
\asix{\phi \psi}{bc}\Big]_{ij} +2  
\mathrm{i}\Big[Y_{a} \asix{\phi  
\psi}{bc} m_\psi^\dagger Y_{c}\Big]_{ij}  
\nonumber\\&\phantom{\sum_{\mathrm{perm}}\Big\{}
-2 \mathrm{i}\Big[Y_{c}  
m_\psi^\dagger Y_{b} \asix{\phi  
\psi}{ac}\Big]_{ij} -1  
\mathrm{i}\mathrm{Tr}\Big[\asix{\phi  
\psi}{ac} Y^\dagger_{b} m_\psi\Big]Y_{ijc}  
+\mathrm{i}\Big[m_\psi Y^\dagger_{a} Y_{c}  
\asix{\phi \psi}{bc}\Big]_{ij}  
\nonumber\\&\phantom{\sum_{\mathrm{perm}}\Big\{}
+\mathrm{i}\Big[m_ 
\psi Y^\dagger_{c} \asix{\phi  
\psi}{ac}^T Y_{b}\Big]_{ij}  
+\mathrm{i}\Big[m_\psi Y^\dagger_{c} Y_{a}  
\asix{\phi \psi}{bc}\Big]_{ij} +3  
\mathrm{i}\Big[Y_{a} Y^\dagger_{c} m_\psi  
\asix{\phi \psi}{bc}\Big]_{ij}  
\nonumber\\&\phantom{\sum_{\mathrm{perm}}\Big\{}
-2 \mathrm{i}\Big[Y_{c}  
Y^\dagger_{b} m_\psi \asix{\phi  
\psi}{ac}\Big]_{ij} 
-2  
\mathrm{i}\kappa_{bcd}\Big[Y_{d}  
\asix{\phi \psi}{ac}\Big]_{ij}
+\frac{3}{2}  
g^2\mathrm{i}\theta^{A}_{bd}\theta^{A}_{cd} 
\Big[m_\psi \asix{\phi  
\psi}{ac}\Big]_{ij}   
\nonumber\\&\phantom{\sum_{\mathrm{perm}}\Big\{}
+\frac{3}{2}  
g^2\mathrm{i}\theta^{A}_{bc}\Big[m_\psi  
\asix{\phi \psi}{ac} t^{A}\Big]_{ij}  
+\frac{9}{2}  
g^2\mathrm{i}\theta^{A}_{bc}\Big[m_\psi t^{A}  
\asix{\phi \psi}{ac}\Big]_{ij}  
-\frac{1}{2} \asix{\bar{\psi}  
\psi}{kilj}Y_{klc}\kappa_{abc}  
\nonumber\\&\phantom{\sum_{\mathrm{perm}}\Big\{}
+2 \asix{\bar{\psi}  
\psi}{kjml}Y_{ikb}\Big[m_\psi^\dagger  
Y_{a}\Big]_{lm} -2 \asix{\bar{\psi}  
\psi}{kilj}\Big[Y_{a} m_\psi^\dagger  
Y_{b}\Big]_{lk} -2 \asix{\bar{\psi}  
\psi}{kjml}Y_{ikb}\Big[Y^\dagger_{a}  
m_\psi\Big]_{lm} \nonumber\\&\phantom{\sum_{\mathrm{perm}}\Big\{}
-4  
\asix{\bar{\psi} \psi}{kilj}\Big[Y_{a}  
Y^\dagger_{b} m_\psi\Big]_{kl} -3  
g\theta^{A}_{bc}\Big[m_\psi Y^\dagger_{c}  
\asix{\psi F}{Aa}\Big]_{ij} +15  
g\Big[m_\psi t^{A} Y^\dagger_{b}  
\asix{\psi F}{Aa}\Big]_{ij}  
\nonumber\\&\phantom{\sum_{\mathrm{perm}}\Big\{}
+3 g\Big[m_\psi  
Y^\dagger_{a} (t^T)^{A} \asix{\psi  
F}{Ab}\Big]_{ij} +12 g\Big[\asix{\psi F}{Aa}  
m_\psi^\dagger Y_{b} t^{A}\Big]_{ij}  
+\frac{1}{2} \asix{\psi  
\phi}{ijbcd}\kappa_{acd}  
\nonumber\\&\phantom{\sum_{\mathrm{perm}}\Big\{}
+\frac{1}{2}  
\Big[m_\psi Y^\dagger_{c} \asix{\psi  
\phi}{abc}\Big]_{ij} +\Big[\asix{\psi  
\phi}{abc} m_\psi^\dagger Y_{c}\Big]_{ij}  
+\frac{1}{4} \asix{\psi  
\psi}{kilj}Y^\dagger_{klc}\kappa_{abc}  
\nonumber\\&\phantom{\sum_{\mathrm{perm}}\Big\{}
-\frac{1}{2}  
\asix{\psi  
\psi}{ijkl}\Big[Y^\dagger_{a} m_\psi  
Y^\dagger_{b}\Big]_{kl} - \asix{\psi  
\psi}{ijkl}\Big[Y^\dagger_{a} Y_{b}  
m_\psi^\dagger\Big]_{kl} \Big\}
,\\
\asixd{\phi \psi}{ijab} =&+\frac{2}{3} g^2\mathrm{i}\Big[
\afive{\phi  
F}{ABc}\afive{\phi F}{CBc}
+\afive{\phi  
\widetilde{F}}{ABc}\afive{\phi  
\widetilde{F}}{CBc}
\Big]
\theta^{A}_{ab}t^{C}_{ij}    
\nonumber\\&
-6 g^2\Big[\afive{\phi  
F}{ABa}\afive{\phi  
\widetilde{F}}{ACb}
+\afive{\phi F}{ACa}\afive{\phi  
\widetilde{F}}{ABb}
\Big]\Big[t^{B} t^{C}\Big]_{ij}   
\nonumber\\&
+g\mathrm{i}\Big[\afive{ 
\phi F}{ABc}-\mathrm{i}\afive{\phi  
\widetilde{F}}{ABc}\Big]\theta^{A}_{ab}\Big[Y^\dagger_{c} 
\afive{\psi F}{B}\Big]_{ij} 
\nonumber\\&
+6 g\mathrm{i}
\afive{\phi  
F}{ABb} 
\theta^{A}_{ac}\Big[Y^\dagger_{c}  
\afive{\psi F}{B}\Big]_{ij}  
+6  
g\afive{\phi \widetilde{F}}{ABa}\Big[t^{A}  
Y^\dagger_{b} \afive{\psi F}{B}\Big]_{ij}  
\nonumber\\&-2 g\afive{\phi  
\widetilde{F}}{ABb}\Big[Y^\dagger_{a}  
\afive{\psi F}{A} t^{B}\Big]_{ij} -8  
g\mathrm{i}\afive{\phi  
F}{ABb}\Big[Y^\dagger_{a} \afive{\psi F}{A} 
t^{B}\Big]_{ij} \nonumber\\&+6  
g\mathrm{i}\afive{\phi  
F}{ABb}\theta^{A}_{ac}\Big[\afive{\psi  
F}{B}^\dagger Y_{c}\Big]_{ij} 
+8  
g\mathrm{i}\afive{\phi  
F}{ABa}\Big[\afive{\psi F}{A}^\dagger  
(t^T)^{B} Y_{b}\Big]_{ij}
\nonumber\\& 
+g\mathrm{i}\Big[\afive{\phi  
F}{ABc}
+\mathrm{i}\afive{\phi  
\widetilde{F}}{ABc}
\Big]\theta^{A}_{ab}\Big[\afive{\psi 
F}{B}^\dagger Y_{c}\Big]_{ij}  
\nonumber\\&-2  
g\afive{\phi  
\widetilde{F}}{ABa}\Big[\afive{\psi  
F}{A}^\dagger (t^T)^{B} Y_{b}\Big]_{ij} +6  
g\afive{\phi  
\widetilde{F}}{ABa}\Big[\afive{\psi  
F}{B}^\dagger Y_{b} t^{A}\Big]_{ij}  
\nonumber\\&+\frac{2}{3}  
g^2\mathrm{i}\theta^{B}_{ab}t^{A}_{ij}\mathrm{Tr} 
\Big[\afive{\psi F}{A} \afive{\psi  
F}{B}^\dagger\Big] +\frac{2}{3}  
g^2\mathrm{i}\theta^{A}_{ab}t^{B}_{ij}\mathrm{Tr} 
\Big[\afive{\psi F}{A} \afive{\psi  
F}{B}^\dagger\Big] 
\nonumber\\&+4  
g^2\mathrm{i}\theta^{A}_{ab}\Big[\afive{ 
\psi F}{B}^\dagger \afive{\psi F}{A}  
t^{B}\Big]_{ij} 
-\frac{8}{3}  
g^2\mathrm{i}\theta^{B}_{ab}\Big[\afive{ 
\psi F}{A}^\dagger \afive{\psi  
F}{A} t^B\Big]_{ij} \nonumber\\&+4  
g^2\mathrm{i}\theta^{B}_{ab}\Big[t^{A}  
\afive{\psi F}{B}^\dagger \afive{\psi  
F}{A}\Big]_{ij} +3  
g^2f_{ABC}\theta^{A}_{ab}\Big[\afive{\psi  
F}{C}^\dagger \afive{\psi F}{B}\Big]_{ij}  
\nonumber\\&-2  
\mathrm{i}\Big[\afive{\psi  
F}{A}^\dagger \afive{\psi F}{A} Y^\dagger_{a}  
Y_{b}\Big]_{ij} 
-2  
\mathrm{i}\Big[Y^\dagger_{a} Y_{b}  
\afive{\psi F}{A}^\dagger \afive{\psi  
F}{A}\Big]_{ij} \nonumber\\&
+3  
g\mathrm{i}\theta^{A}_{bc}\Big[\afive{ 
\psi \phi^2 }{ac}^\dagger \afive{\psi  
F}{A}\Big]_{ij} +3  
g\mathrm{i}\theta^{A}_{bc}\Big[\afive{ 
\psi F}{A}^\dagger \afive{\psi \phi^2  
}{ac}\Big]_{ij} \nonumber\\&
-\frac{1}{2}  
\mathrm{i}\Big[\afive{\psi \phi^2  
}{ac}^\dagger \afive{\psi \phi^2  
}{bc}\Big]_{ij}  
-\frac{1}{2}  
g^3\mathrm{i}\asix{3F}{ABC}f_{DBC}\Big[\theta^{D}_{ab}t^{ 
A}_{ij}  + \theta^{A}_{ab}t^{ 
D}_{ij} \Big]
\nonumber\\&+g^3\asix{3F}{ABC} 
\theta^{A}_{ab}\Big[t^{B} t^{C}\Big]_{ij} -4  
\mathrm{i}\asix{\phi  
D}{acbd}\Big[Y^\dagger_{d} Y_{c}\Big]_{ij}  
\nonumber\\& -\frac{1}{6} g^2\asix{\phi  
\psi}{ijcd}\theta^{A}_{ab}\theta^{A}_{cd}  
-\frac{1}{2}  
\asix{\phi  
\psi}{ijbc}\mathrm{Tr}\Big[Y_{a}  
Y^\dagger_{c} + Y_c Y_a^\dagger\Big]   
\nonumber\\&+\frac{1}{4}  
\Big[\asix{\phi \psi}{ab} Y^\dagger_{c}  
Y_{c}\Big]_{ij} +\Big[\asix{\phi  
\psi}{ac} Y^\dagger_{c} Y_{b}\Big]_{ij}  
\nonumber\\&+2 \Big[\asix{\phi  
\psi}{ac} Y^\dagger_{b} Y_{c}\Big]_{ij} -  
\Big[Y^\dagger_{a} Y_{c} \asix{\phi  
\psi}{bc}\Big]_{ij}  
\nonumber\\&+\frac{1}{2}  
\Big[Y^\dagger_{c} \asix{\phi \psi}{ab}^T  
Y_{c}\Big]_{ij} -2 \Big[Y^\dagger_{c} Y_{a} 
\asix{\phi \psi}{bc}\Big]_{ij}  
\nonumber\\&+\frac{1}{4}  
\Big[Y^\dagger_{c} Y_{c} \asix{\phi  
\psi}{ab}\Big]_{ij} -4  
g^2\theta^{A}_{bc}\Big[\asix{\phi  
\psi}{ac} t^{A}\Big]_{ij}  
\nonumber\\&+4  
g^2\theta^{A}_{bc}\Big[t^{A} \asix{\phi  
\psi}{ac}\Big]_{ij} +\frac{1}{2}  
g^2\Big[\asix{\phi \psi}{ab} t^{A}  
t^{A}\Big]_{ij} \nonumber\\&-  
g^2\Big[t^{A} \asix{\phi \psi}{ab}  
t^{A}\Big]_{ij} +\frac{1}{2} g^2\Big[t^{A} t^{A}  
\asix{\phi \psi}{ab}\Big]_{ij}  
\nonumber\\&+\frac{2}{3}  
g^2\mathrm{i}\asix{\bar{\psi}  
\psi}{ijkl}\theta^{A}_{ab}t^{A}_{lk} -2  
\mathrm{i}\asix{\bar{\psi}  
\psi}{ijkl}\Big[Y_{a} Y^\dagger_{b}\Big]_{kl}  
\nonumber\\&+3  
g\mathrm{i}\theta^{A}_{ac}\Big[Y^\dagger_{c}
\asix{\psi F}{Ab}\Big]_{ij} -3  
g\mathrm{i}\Big[t^{A} Y^\dagger_{b}  
\asix{\psi F}{Aa}\Big]_{ij}  
\nonumber\\&+3  
g\mathrm{i}\Big[Y^\dagger_{a} (t^T)^{A}  
\asix{\psi F}{Ab}\Big]_{ij} 
- (a\leftrightarrow b)
,\\
\asixd{\bar{\psi} \psi}{ijkl} =&
+\frac{1}{3} g^2 \Big[
\afive{\phi  
F}{ABa}\afive{\phi F}{CBa} +\afive{\phi  
\widetilde{F}}{ABa}\afive{\phi  
\widetilde{F}}{CBa}\Big]t^{A}_{kl}t^{C}_{ij}    
\nonumber\\&
-\frac{1}{4} \Big[ 
\afive{\phi F}{ABa}\afive{\phi  
F}{ABb} +\afive{\phi  
\widetilde{F}}{ABa}\afive{\phi  
\widetilde{F}}{ABb}
\Big]Y_{jla}Y^\dagger_{ikb}   
\nonumber\\&
-2 g\Big[
\afive{\phi  
F}{ABa}
-\mathrm{i}\afive{\phi  
\widetilde{F}}{ABa}
\Big]Y^\dagger_{ika}\Big[\afive{\psi F}{A}  
t^{B}\Big]_{jl} 
\nonumber\\&
+2 g\Big[
\afive{\phi  
F}{ABa}-\mathrm{i}\afive{\phi  
\widetilde{F}}{ABa}
\Big]t^{A}_{kj}\Big[Y^\dagger_{a}  
\afive{\psi F}{B}\Big]_{il}   
\nonumber\\&
+2 g\Big[\afive{\phi  
F}{ABa} +\mathrm{i}\afive{\phi  
\widetilde{F}}{ABa}
\Big]Y_{jla}\Big[\afive{\psi F}{A}^\dagger  
(t^T)^{B}\Big]_{ki} 
\nonumber\\&
+2 g\Big[\afive{\phi  
F}{ABa} +\mathrm{i}\afive{\phi  
\widetilde{F}}{ABa}
\Big]t^{A}_{ij}\Big[\afive{\psi  
F}{B}^\dagger Y_{a}\Big]_{kl}    
\nonumber\\&+3 \Big[Y^\dagger_{a}  
\afive{\psi  
F}{A}\Big]_{ij}\Big[Y^\dagger_{a}  
\afive{\psi F}{A}\Big]_{kl} -3  
Y^\dagger_{ika}\Big[\afive{\psi F}{A}  
Y^\dagger_{a} \afive{\psi F}{A}\Big]_{jl}  
\nonumber\\&
-2  
g^2\mathrm{i}f_{ABC}t^{A}_{ij}\Big[\afive{ 
\psi F}{B}^\dagger \afive{\psi  
F}{C}\Big]_{kl} 
\nonumber\\&
-\frac{16}{3}  
g^2t^{B}_{ij}\Big[\afive{\psi F}{A}^\dagger 
\afive{\psi F}{A} t^B\Big]_{kl} 
+6 g^2\Big[\afive{\psi F}{B}^\dagger  
\afive{\psi F}{A}\Big]_{ij}\Big[t^{A}  
t^{B}\Big]_{kl} \nonumber\\&+18  
g^2\Big[\afive{\psi F}{B}^\dagger  
\afive{\psi F}{A}\Big]_{ij}\Big[t^{B}  
t^{A}\Big]_{kl} -12 g^2\Big[\afive{\psi F}{A}  
t^{B}\Big]_{jl}\Big[t^{A} \afive{\psi  
F}{B}^\dagger\Big]_{ik} \nonumber\\&-12  
g^2\Big[\afive{\psi F}{A}  
t^{B}\Big]_{jl}\Big[t^{B} \afive{\psi  
F}{A}^\dagger\Big]_{ik} +8 g^2t^{B}_{ij}\Big[t^{A}  
\afive{\psi F}{B}^\dagger \afive{\psi  
F}{A}\Big]_{kl} \nonumber\\&+8  
g^2t^{A}_{ij}\Big[\afive{\psi F}{B}^\dagger  
\afive{\psi F}{A} t^{B}\Big]_{kl}  
+\frac{2}{3}  
g^2t^{A}_{ij}t^{B}_{kl}\mathrm{Tr}\Big[\afive{ 
\psi F}{A} \afive{\psi  
F}{B}^\dagger\Big] \nonumber\\&+6  
\Big[\afive{\psi F}{A}^\dagger  
Y_{a}\Big]_{kl}\Big[Y^\dagger_{a}  
\afive{\psi F}{A}\Big]_{ij} -3  
Y_{jla}\Big[\afive{\psi F}{A}^\dagger Y_{a}  
\afive{\psi F}{A}^\dagger\Big]_{ki}  
\nonumber\\&+3 \Big[\afive{\psi  
F}{A}^\dagger  
Y_{a}\Big]_{ij}\Big[\afive{\psi  
F}{A}^\dagger Y_{a}\Big]_{kl} 
-\frac{1}{8}  
\afive{\psi \phi^2 }{jlab}\afive{\psi  
\phi^2 }{ikab}^\dagger  
\nonumber\\&
+\frac{1}{2}  
g^3\asix{3F}{ABC}f_{DBC}t^{A}_{ij}t^{D}_{kl} 
+\frac{1}{3}  
g^2\mathrm{i}\asix{\phi  
\psi}{klab}\theta^{A}_{ab}t^{A}_{ij}  
+\mathrm{i}\asix{\phi  
\psi}{klab}\Big[Y^\dagger_{a} Y_{b}\Big]_{ij}  
\nonumber\\& +2  
g^2\asix{\bar{\psi}  
\psi}{mjnl}t^{A}_{in}t^{A}_{km}  
-2  
g^2\asix{\bar{\psi}  
\psi}{ilmn}t^{A}_{km}t^{A}_{nj} +2  
g^2\asix{\bar{\psi}  
\psi}{imkn}t^{A}_{ml}t^{A}_{nj}  
\nonumber\\&+\frac{1}{2}  
g^2\asix{\bar{\psi}  
\psi}{ijkm}\Big[t^{A} t^{A}\Big]_{ml}  
+\frac{1}{2} g^2\asix{\bar{\psi}  
\psi}{ijml}\Big[t^{A} t^{A}\Big]_{km}  
\nonumber\\&+\frac{4}{3}  
g^2\asix{\bar{\psi}  
\psi}{klmn}t^{A}_{ij}t^{A}_{nm} +\frac{1}{2}  
\asix{\bar{\psi}  
\psi}{mjnl}Y_{mna}Y^\dagger_{ika}  
\nonumber\\&+\frac{1}{2}  
\asix{\bar{\psi}  
\psi}{klmn}Y_{jma}Y^\dagger_{ina} +\frac{1}{4}  
\asix{\bar{\psi} \psi}{mjkl}\Big[Y_{a}  
Y^\dagger_{a}\Big]_{mi}  
\nonumber\\&+\frac{1}{2}  
\asix{\bar{\psi}  
\psi}{inkm}Y_{jla}Y^\dagger_{mna} +\frac{1}{4}  
\asix{\bar{\psi} \psi}{imkl}\Big[Y_{a}  
Y^\dagger_{a}\Big]_{jm}  
\nonumber\\&+\frac{1}{2}  
\asix{\bar{\psi}  
\psi}{inml}Y_{jma}Y^\dagger_{kna} -\frac{1}{12}  
\asix{\psi  
\psi}{mlnj}Y^\dagger_{ina}Y^\dagger_{kma}  
\nonumber\\&+\frac{1}{12}  
\asix{\psi  
\psi}{jmln}Y^\dagger_{ina}Y^\dagger_{kma}  
-\frac{1}{4} \asix{\psi  
\psi}{mjnl}Y^\dagger_{ika}Y^\dagger_{mna}  
\nonumber\\&-\frac{1}{4}  
\overline{\asix{\psi  
\psi}{nimk}}Y_{jla}Y_{mna} -\frac{1}{12}  
\overline{\asix{\psi  
\psi}{nimk}}Y_{jna}Y_{lma}  
\nonumber\\&+\frac{1}{12}  
\overline{\asix{\psi  
\psi}{imkn}}Y_{jna}Y_{lma}
+(i\leftrightarrow k)
+(j\leftrightarrow l)
+(i\leftrightarrow k)(j\leftrightarrow l)
,\\
\asixd{\psi F}{Aija} =&
+2 g\Big[
\afive{\phi F}{ABb}\afive{\phi  
F}{BCa}+\mathrm{i}\afive{\phi  
F}{BCa}\afive{\phi  
\widetilde{F}}{ABb}
\Big]
\Big[Y_{b} t^{C}\Big]_{ij}   
\nonumber\\&
-2 
g^2\Big[\afive{\phi F}{ABa}+\mathrm{i}\afive{\phi  
\widetilde{F}}{ABa}\Big]\afive{\psi  
F}{Cij}f_{BDE}f_{CDE} 
\nonumber\\&
-5 g^2\afive{\phi  
F}{BCa}\afive{\psi F}{Dij}f_{ACE}f_{BDE} 
-2  
g^2\mathrm{i}\afive{\phi  
\widetilde{F}}{BCa}\afive{\psi  
F}{Dij}f_{ACE}f_{BDE} 
\nonumber\\&-  
\mathrm{i}g^2\Big[\afive{\phi  
F}{ABb} + \mathrm{i}\afive{\phi  
\widetilde{F}}{ABb}
\Big]\afive{\psi F}{Cij}f_{BCD}\theta^{D}_{ab}  
\nonumber\\&
-2  
g^2\Big[\afive{\phi  
F}{ABb}+\mathrm{i}\afive{\phi 
\widetilde{F}}{ABb}\Big]\theta^{C}_{ab}\Big[\afive{\psi F}{B}  
t^{C}\Big]_{ij} 
\nonumber\\& 
-6 g^2\mathrm{i}\Big[\afive{\phi  
F}{ABa} +\mathrm{i}\afive{\phi  
\widetilde{F}}{ABa}
\Big]f_{BCD}\Big[\afive{\psi F}{C}  
t^{D}\Big]_{ij}
\nonumber\\&
-4 g^2\afive{\phi  
\widetilde{F}}{BCa}f_{ACD}\Big[\afive{\psi  
F}{B} t^{D}\Big]_{ij} 
+18 g^2\mathrm{i}\afive{\phi  
F}{BCa}f_{ACD}\Big[\afive{\psi F}{D}  
t^{B}\Big]_{ij} \nonumber\\&-6  
g^2\mathrm{i}\afive{\phi  
\widetilde{F}}{BCa}\Big[\afive{\psi F}{A}  
t^{C} t^{B}\Big]_{ij} -2  
g^2\mathrm{i}\afive{\phi  
\widetilde{F}}{BCa}\Big[\afive{\psi F}{B}  
t^{C} t^{A}\Big]_{ij} \nonumber\\&
+8 g^2\afive{\phi  
F}{BCa}\Big[\afive{\psi F}{B} t^{C}  
t^{A}\Big]_{ij} \nonumber\\&
+2 g^2\Big[\afive{\phi F}{BCa}+\mathrm{i}\afive{\phi  
\widetilde{F}}{BCa}\Big]\Big[\afive{\psi  
F}{B} t^{A} t^{C}\Big]_{ij} 
\nonumber\\&
+2 \Big[\afive{\phi  
F}{ABb}+\mathrm{i}\afive{\phi  
\widetilde{F}}{ABb}\Big]\Big[\afive{\psi F}{B} Y^\dagger_{a} 
Y_{b}\Big]_{ij}  
\nonumber\\&
+\Big[\afive{\phi  
F}{ABb}
+\mathrm{i}\afive{ 
\phi  
\widetilde{F}}{ABb}\Big]\Big[\afive{\psi F}{B} Y^\dagger_{b} 
Y_{a}\Big]_{ij} 
\nonumber\\&
+\Big[\afive{\phi  
F}{ABb}
+\mathrm{i}\afive{ 
\phi \widetilde{F}}{ABb}
\Big]\Big[Y_{a} \afive{\psi F}{B}^\dagger 
Y_{b}\Big]_{ij}  
\nonumber\\&
- g\Big[\afive{\phi F}{ABb}+\mathrm{i}\afive{\phi  
\widetilde{F}}{ABb}\Big]
\Big[\afive{\psi  
\phi^2 }{ab} t^{B}\Big]_{ij}  
+6 g\theta^{B}_{ab}\Big[\afive{\psi F}{A}  
Y^\dagger_{b} \afive{\psi F}{B}\Big]_{ij}  
\nonumber\\&
+4  
g\mathrm{i}f_{ABC}\Big[\afive{\psi F}{B}  
Y^\dagger_{a} \afive{\psi F}{C}\Big]_{ij} +6  
g\Big[\afive{\psi F}{B} t^{A} Y^\dagger_{a}  
\afive{\psi F}{B}\Big]_{ij}  
\nonumber\\&-6  
g\theta^{B}_{ab}\Big[\afive{\psi F}{A}  
\afive{\psi F}{B}^\dagger Y_{b}\Big]_{ij} -4  
g\mathrm{i}f_{ACB}\Big[\afive{\psi F}{B}  
\afive{\psi F}{C}^\dagger Y_{a}\Big]_{ij}  
\nonumber\\&+6 g\Big[\afive{\psi  
F}{A} \afive{\psi F}{B}^\dagger Y_{a}  
t^{B}\Big]_{ij} +2 g\Big[\afive{\psi F}{B}  
\afive{\psi F}{B}^\dagger Y_{a}  
t^{A}\Big]_{ij} \nonumber\\&+2  
g\Big[\afive{\psi F}{B} t^{A}  
\afive{\psi F}{B}^\dagger Y_{a}\Big]_{ij} -2  
g\Big[Y_{a} \afive{\psi F}{B}^\dagger  
\afive{\psi F}{A} t^{B}\Big]_{ij}  
\nonumber\\&
-\Big[\afive{\psi  
F}{A} Y^\dagger_{b} \afive{\psi \phi^2  
}{ab}\Big]_{ij} +\frac{1}{2}  
g^2\mathrm{i}\Big[\asix{3F}{ABC}+\mathrm{i}\asix{3\widetilde{F}}{ABC}\Big]\Big[(t^T)^{B} Y_{a}  
t^{C}\Big]_{ij}  
\nonumber\\&
+\Big[\asix{\phi  
F}{ABab}+\mathrm{i}\asix{\phi  
\widetilde{F}}{ABab}\Big]\Big[Y_{b} t^{B}\Big]_{ij}  
\nonumber\\&
-\mathrm{i} g\Big[\asix{\phi \psi}{ab}^T  
Y_{b} t^{A}\Big]_{ij} +\frac{1}{6}  
g^2\asix{\psi F}{Bija}f_{ACD}f_{BCD}  
\nonumber\\&
- g^2\asix{\psi  
F}{Bijb}\Big[\theta^{A}_{bc}\theta^{B}_{ac}+\theta^{A}_{ac}\theta^{B}_{bc} \Big] 
-\frac{1}{12}  
g^2\asix{\psi  
F}{Bija}\theta^{A}_{bc}\theta^{B}_{bc}  
\nonumber\\&
+g^2\asix{\psi  
F}{Aijb}\theta^{B}_{ac}\theta^{B}_{cb}  
+\frac{1}{3}  
g^2\asix{\psi F}{Bija}\mathrm{Tr}\Big[t^{B} 
t^{A}\Big] +g^2\Big[\asix{\psi F}{Aa} t^{B} 
t^{B}\Big]_{ij} \nonumber\\&+3  
g^2\Big[\asix{\psi F}{Ba} t^{B}  
t^{A}\Big]_{ij} +g^2\Big[\asix{\psi F}{Ba}  
t^{A} t^{B}\Big]_{ij}  
\nonumber\\&
+\frac{1}{4}  
\asix{\psi F}{Aijb}\mathrm{Tr}\Big[Y_{a}  
Y^\dagger_{b}+Y_{b}  
Y^\dagger_{a}\Big]  
\nonumber\\&+\frac{1}{2}  
\Big[\asix{\psi F}{Aa} Y^\dagger_{b}  
Y_{b}\Big]_{ij} +\Big[\asix{\psi F}{Ab}  
Y^\dagger_{b} Y_{a}\Big]_{ij}  
\nonumber\\&+2 \Big[\asix{\psi  
F}{Ab} Y^\dagger_{a} Y_{b}\Big]_{ij} -2  
g^2\theta^{B}_{ab}\Big[\asix{\psi F}{Ab}  
t^{B}\Big]_{ij} \nonumber\\&+7  
\mathrm{i}g^2f_{ABC}\Big[\asix{\psi F}{Ba}  
t^{C}\Big]_{ij} +\frac{1}{6} g\asix{\psi  
\psi}{kijl}\Big[t^{A} Y^\dagger_{a}\Big]_{lk}  
\nonumber\\&
-(i\leftrightarrow j)
,\\
\asixd{\psi \phi}{ijabc} =&\sum_{\pi(ij),\pi(abc)}\Big\{
-\frac{1}{12}  
\afive{\phi}{abcde}\afive{\psi \phi^2  
}{ijde} \nonumber\\&
\phantom{\sum_{\pi(ij),\pi(abc)}}
+2 g^2\Big[\afive{\phi  
F}{ABa}\afive{\phi  
F}{ACd}+\afive{\phi  
\widetilde{F}}{ABa}\afive{\phi  
\widetilde{F}}{ACd}\Big]
\theta^{B}_{ce}\theta^{C}_{be}Y_{ijd} 
\nonumber\\&
\phantom{\sum_{\pi(ij),\pi(abc)}}
- g^2\afive{\phi  
\widetilde{F}}{ABa}\afive{\phi  
\widetilde{F}}{ACb}\theta^{B}_{cd}\theta^{C}_{ed}Y_{ 
ije}
+5  
g^2\afive{\phi F}{ABa}\afive{\phi  
F}{ACb}\theta^{B}_{cd}\theta^{C}_{ed}Y_{ije} 
\nonumber\\&\phantom{\sum_{\pi(ij),\pi(abc)}}
-\frac{1}{6}  \Big[
\afive{\phi F}{ABd}\afive{\phi  
F}{ABe}+\afive{\phi  
\widetilde{F}}{ABd}\afive{\phi  
\widetilde{F}}{ABe}
\Big]Y_{ijd}\lambda_{abce}  
\nonumber\\&\phantom{\sum_{\pi(ij),\pi(abc)}}
+6  
g^2\mathrm{i}\afive{\phi  
F}{ABb}\afive{\phi  
\widetilde{F}}{ACc}\Big[Y_{a} t^{B} t^{C}\Big]_{ij}  
+6 g^2\mathrm{i}\afive{\phi  
F}{ABb}\afive{\phi  
\widetilde{F}}{ACc}\Big[Y_{a} t^{C} t^{B}\Big]_{ij}  
\nonumber\\&\phantom{\sum_{\pi(ij),\pi(abc)}}
-12 g^2\afive{\phi  
F}{ABc}\afive{\phi F}{BCb}\Big[(t^T)^{A} Y_{a}  
t^{C}\Big]_{ij} -24 g^3\afive{\phi  
F}{ABa}\theta^{B}_{cd}\theta^{C}_{bd}\Big[ 
\afive{\psi F}{A} t^{C}\Big]_{ij}  
\nonumber\\&\phantom{\sum_{\pi(ij),\pi(abc)}}
-24 g^3\afive{\phi  
F}{ABa}\theta^{B}_{cd}\theta^{C}_{bd}\Big[ 
\afive{\psi F}{C} t^{A}\Big]_{ij}  
\nonumber\\&\phantom{\sum_{\pi(ij),\pi(abc)}}
-\frac{2}{3} g\Big[\afive{\phi  
F}{ABd}-\mathrm{i}\afive{\phi  
\widetilde{F}}{ABd}\Big]\lambda_{abcd}\Big[\afive{\psi F}{A} 
t^{B}\Big]_{ij}  
+6 g\afive{\phi  
F}{ABa}\theta^{B}_{bd}\Big[\afive{\psi F}{A}  
Y^\dagger_{d} Y_{c}\Big]_{ij}  
\nonumber\\&\phantom{\sum_{\pi(ij),\pi(abc)}}
+24 g\afive{\phi  
F}{ABc}\Big[\afive{\psi F}{A} Y^\dagger_{a} 
Y_{b} t^{B}\Big]_{ij} -8 g\afive{\phi  
F}{ABa}\Big[\afive{\psi F}{A} t^{B}  
Y^\dagger_{b} Y_{c}\Big]_{ij}  
\nonumber\\&\phantom{\sum_{\pi(ij),\pi(abc)}}
-6  
g\mathrm{i}\afive{\phi  
\widetilde{F}}{ABa}\Big[\afive{\psi F}{A}  
Y^\dagger_{b} (t^T)^{B} Y_{c}\Big]_{ij} +2  
g\mathrm{i}\afive{\phi  
\widetilde{F}}{ABa}\Big[\afive{\psi F}{A}  
t^{B} Y^\dagger_{b} Y_{c}\Big]_{ij}  
\nonumber\\&
\phantom{\sum_{\pi(ij),\pi(abc)}}
-2 g\afive{\phi  
F}{ABb}\theta^{B}_{cd}\Big[Y_{a} \afive{\psi  
F}{A}^\dagger Y_{d}\Big]_{ij} +8 g\afive{\phi  
F}{ABb}\Big[Y_{a} \afive{\psi F}{A}^\dagger  
Y_{c} t^{B}\Big]_{ij} \nonumber\\&
\phantom{\sum_{\pi(ij),\pi(abc)}}
+6  
g\mathrm{i}\afive{\phi  
\widetilde{F}}{ABb}\Big[Y_{a} \afive{\psi  
F}{A}^\dagger Y_{c} t^{B}\Big]_{ij} -2  
g\mathrm{i}\afive{\phi  
\widetilde{F}}{ABb}\Big[Y_{a} \afive{\psi  
F}{A}^\dagger (t^T)^{B} Y_{c}\Big]_{ij}  
\nonumber\\&\phantom{\sum_{\pi(ij),\pi(abc)}}
-3 g^2\afive{\phi  
F}{ABb}\afive{\psi \phi^2  
}{ijcd}\theta^{A}_{de}\theta^{B}_{ea} -3  
g^2\mathrm{i}\afive{\phi  
\widetilde{F}}{ABc}\Big[\afive{\psi  
\phi^2 }{ab} t^{B} t^{A}\Big]_{ij}  
\nonumber\\&\phantom{\sum_{\pi(ij),\pi(abc)}}
+3 g^2\afive{\phi  
F}{ABc}\Big[(t^T)^{A} \afive{\psi \phi^2  
}{ab} t^{B}\Big]_{ij} -12  
g^2\theta^{A}_{cd}\theta^{B}_{bd}\Big[\afive{ 
\psi F}{A} Y^\dagger_{a} \afive{\psi  
F}{B}\Big]_{ij} \nonumber\\&
\phantom{\sum_{\pi(ij),\pi(abc)}}
- \lambda_{abcd}\Big[\afive{\psi F}{A}  
Y^\dagger_{d} \afive{\psi F}{A}\Big]_{ij} +12  
\Big[\afive{\psi F}{A} Y^\dagger_{a} Y_{b}  
Y^\dagger_{c} \afive{\psi F}{A}\Big]_{ij}  
\nonumber\\&\phantom{\sum_{\pi(ij),\pi(abc)}}
-6  
g^2\theta^{A}_{bd}\theta^{B}_{ad}\Big[\afive{ 
\psi F}{A} \afive{\psi F}{B}^\dagger  
Y_{c}\Big]_{ij} +12 \Big[\afive{\psi F}{A}  
Y^\dagger_{b} Y_{a} \afive{\psi F}{A}^\dagger  
Y_{c}\Big]_{ij} \nonumber\\&
\phantom{\sum_{\pi(ij),\pi(abc)}}
-2  
\Big[\afive{\psi F}{A} \afive{\psi  
F}{A}^\dagger Y_{a} Y^\dagger_{b} Y_{c}\Big]_{ij}  
-2 \Big[Y_{a} \afive{\psi F}{A}^\dagger  
\afive{\psi F}{A} Y^\dagger_{b}  
Y_{c}\Big]_{ij} \nonumber\\&
\phantom{\sum_{\pi(ij),\pi(abc)}}
-3  g\theta^{A}_{ad}\Big[\afive{\psi F}{A}  
\afive{\psi \phi^2 }{bd}^\dagger  
Y_{c}\Big]_{ij} -6 g\Big[\afive{\psi F}{A}  
\afive{\psi \phi^2 }{bc}^\dagger Y_{a}  
t^{A}\Big]_{ij} \nonumber\\&\phantom{\sum_{\pi(ij),\pi(abc)}}
-3  
g\theta^{A}_{ad}\Big[\afive{\psi F}{A}  
Y^\dagger_{d} \afive{\psi \phi^2  
}{bc}\Big]_{ij} -6 g\Big[\afive{\psi F}{A}  
Y^\dagger_{a} \afive{\psi \phi^2 }{bc}  
t^{A}\Big]_{ij} \nonumber\\&
\phantom{\sum_{\pi(ij),\pi(abc)}}
-3  
g\theta^{A}_{cd}\Big[\afive{\psi \phi^2  
}{ab} \afive{\psi F}{A}^\dagger  
Y_{d}\Big]_{ij} -3  
g\theta^{A}_{bd}\Big[\afive{\psi \phi^2  
}{ad} \afive{\psi F}{A}^\dagger  
Y_{c}\Big]_{ij} \nonumber\\&
\phantom{\sum_{\pi(ij),\pi(abc)}}
+3  
g\Big[\afive{\psi \phi^2 }{ab}  
\afive{\psi F}{A}^\dagger Y_{c}  
t^{A}\Big]_{ij}  
-3  
g\Big[Y_{a} \afive{\psi F}{A}^\dagger  
\afive{\psi \phi^2 }{bc} t^{A}\Big]_{ij}  
\nonumber\\&\phantom{\sum_{\pi(ij),\pi(abc)}}
-\frac{1}{4} \afive{\psi \phi^2  
}{ijcd}\mathrm{Tr}\Big[\afive{\psi  
\phi^2 }{ab} Y^\dagger_{d}\Big]  
\nonumber\\&\phantom{\sum_{\pi(ij),\pi(abc)}}
-\frac{1}{2}  
\Big[\afive{\psi \phi^2 }{ab}  
Y^\dagger_{d} \afive{\psi \phi^2  
}{cd}\Big]_{ij} - \Big[\afive{\psi  
\phi^2 }{ad} Y^\dagger_{b} \afive{\psi  
\phi^2 }{cd}\Big]_{ij}  
\nonumber\\&\phantom{\sum_{\pi(ij),\pi(abc)}}
+\frac{1}{12}  
\mathrm{Tr}\Big[\afive{\psi \phi^2  
}{ab} \afive{\psi \phi^2  
}{cd}^\dagger\Big]Y_{ijd} +\frac{1}{12}  
\mathrm{Tr}\Big[\afive{\psi \phi^2  
}{ad} \afive{\psi \phi^2  
}{bc}^\dagger\Big]Y_{ijd}  
\nonumber\\&
\phantom{\sum_{\pi(ij),\pi(abc)}}
-\frac{1}{4}  
\afive{\psi \phi^2  
}{ijcd}\mathrm{Tr}\Big[\afive{\psi  
\phi^2 }{ab}^\dagger Y_{d}\Big] -\frac{1}{2}  
\Big[\afive{\psi \phi^2 }{ad}  
\afive{\psi \phi^2 }{bd}^\dagger  
Y_{c}\Big]_{ij} \nonumber\\&\phantom{\sum_{\pi(ij),\pi(abc)}}
-  
\Big[\afive{\psi \phi^2 }{ad}  
\afive{\psi \phi^2 }{bc}^\dagger  
Y_{d}\Big]_{ij} +\frac{3}{2} g^2\asix{\phi  
D}{daef}\theta^{A}_{be}\theta^{A}_{cf}Y_{ijd} 
\nonumber\\&\phantom{\sum_{\pi(ij),\pi(abc)}}
+\frac{3}{2}  
g^2\asix{\phi  
D}{dfbc}\theta^{A}_{ad}\theta^{A}_{ef}Y_{ije} 
-\frac{3}{2} g^2\asix{\phi  
D}{dfab}\theta^{A}_{ce}\theta^{A}_{ef}Y_{ijd} 
\nonumber\\&\phantom{\sum_{\pi(ij),\pi(abc)}}
+\frac{1}{2}  
\asix{\phi D}{dfbc}Y_{ije}\lambda_{aedf} 
-2  
\asix{\phi D}{bdce}\Big[Y_{a} Y^\dagger_{e} 
Y_{d}\Big]_{ij} \nonumber\\&\phantom{\sum_{\pi(ij),\pi(abc)}}
-4  
\asix{\phi D}{bdce}\Big[Y_{d} Y^\dagger_{a} 
Y_{e}\Big]_{ij} 
-\frac{3}{2} g^2\asix{\phi F}{ABbc}\theta^{A}_{de}\theta^{B}_{ea}Y_{ijd} 
\nonumber\\&\phantom{\sum_{\pi(ij),\pi(abc)}}
-3 g^2\asix{\phi  
F}{ABbc}\Big[(t^T)^{A} Y_{a} t^{B}\Big]_{ij} +3  
g^2\mathrm{i}\asix{\phi  
\widetilde{F}}{ABbc}\Big[Y_{a} t^{A}  
t^{B}\Big]_{ij} \nonumber\\&\phantom{\sum_{\pi(ij),\pi(abc)}}
-\frac{3}{2} \mathrm{i}
g^2\theta^{A}_{bd}\theta^{A}_{ce} 
\Big[Y_{a} \asix{\phi \psi  
}{ed}\Big]_{ij}
-6  
\mathrm{i}\lambda_{abde}\Big[Y_{e}  
\asix{\phi \psi}{cd}\Big]_{ij} -6  
g^2\mathrm{i}\theta^{A}_{cd}\Big[\asix{ 
\phi \psi}{ad}^T Y_{b} t^{A}\Big]_{ij}  
\nonumber\\&\phantom{\sum_{\pi(ij),\pi(abc)}}
+\mathrm{i}\Big[Y_{a}  
Y^\dagger_{b} Y_{d} \asix{\phi  
\psi}{cd}\Big]_{ij} +\mathrm{i}\Big[Y_{a}  
Y^\dagger_{d} Y_{b} \asix{\phi  
\psi}{cd}\Big]_{ij} \nonumber\\&
\phantom{\sum_{\pi(ij),\pi(abc)}}
-2  
\mathrm{i}\Big[Y_{d} Y^\dagger_{a} Y_{b}  
\asix{\phi \psi}{cd}\Big]_{ij}  
+\frac{1}{6} \asix{\bar{\psi}  
\psi}{kilj}Y_{kld}\lambda_{abcd}  
\nonumber\\&
\phantom{\sum_{\pi(ij),\pi(abc)}}
-2 \asix{\bar{\psi}  
\psi}{kilj}\Big[Y_{a} Y^\dagger_{b}  
Y_{c}\Big]_{kl} -12  
g^3\theta^{A}_{cd}\theta^{B}_{bd}\Big[\asix{
\psi F}{Aa} t^{B}\Big]_{ij}  
\nonumber\\&\phantom{\sum_{\pi(ij),\pi(abc)}}
+6  
g\theta^{A}_{bd}\Big[\asix{\psi F}{Aa}  
Y^\dagger_{d} Y_{c}\Big]_{ij} +12  
g\Big[\asix{\psi F}{Aa} Y^\dagger_{b} Y_{c} 
t^{A}\Big]_{ij}  
\nonumber\\&\phantom{\sum_{\pi(ij),\pi(abc)}}
+\frac{1}{2}  
g^2\asix{\psi  
\phi}{ijade}\theta^{A}_{bd}\theta^{A}_{ce}  
+\frac{1}{2}  
g^2\asix{\psi  
\phi}{ijabd}\theta^{A}_{ec}\theta^{A}_{ed}  
+\frac{1}{8} \asix{\psi  
\phi}{ijbcd}\mathrm{Tr}\Big[Y_{a}  
Y^\dagger_{d}\Big]  
\nonumber\\&\phantom{\sum_{\pi(ij),\pi(abc)}}
+\frac{1}{8}  
\asix{\psi  
\phi}{ijabd}\mathrm{Tr}\Big[Y_{d}  
Y^\dagger_{c}\Big] 
-\frac{1}{12}  
\mathrm{Tr}\Big[\asix{\psi \phi}{abc}  
Y^\dagger_{d}\Big]Y_{ijd}  
\nonumber\\&\phantom{\sum_{\pi(ij),\pi(abc)}}
+\frac{1}{12}  
\Big[\asix{\psi \phi}{abc}  
Y^\dagger_{d} Y_{d}\Big]_{ij} +\frac{1}{2}  
\Big[\asix{\psi \phi}{abd}  
Y^\dagger_{d} Y_{c}\Big]_{ij}  
\nonumber\\&\phantom{\sum_{\pi(ij),\pi(abc)}}
+\Big[\asix{\psi  
\phi}{abd} Y^\dagger_{c} Y_{d}\Big]_{ij}  
+\frac{1}{4} \asix{\psi  
\phi}{ijcde}\lambda_{abde}  
\nonumber\\&\phantom{\sum_{\pi(ij),\pi(abc)}}
+g^2\theta^{A}_{cd}\Big[ 
\asix{\psi \phi}{abd} t^{A}\Big]_{ij}  
-\frac{1}{2} g^2\Big[\asix{\psi  
\phi}{abc} t^{A} t^{A}\Big]_{ij}  
\nonumber\\&\phantom{\sum_{\pi(ij),\pi(abc)}}
-\frac{1}{12}  
\mathrm{Tr}\Big[\asix{\psi  
\phi}{abc}^\dagger Y_{d}\Big]Y_{ijd}  
+\frac{1}{6} \Big[Y_{d} \asix{\psi  
\phi}{abc}^\dagger Y_{d}\Big]_{ij}  
\nonumber\\&\phantom{\sum_{\pi(ij),\pi(abc)}}
-\frac{1}{12}  
\asix{\psi  
\psi}{kilj}Y^\dagger_{kld}\lambda_{abcd}  
+\asix{\psi  
\psi}{kilj}\Big[Y^\dagger_{a} Y_{b}  
Y^\dagger_{c}\Big]_{kl} \Big\}
,\\
\asixd{\psi \psi}{ijkl} =&+2 \afive{\phi F}{ABa}\afive{\phi  
F}{ABb}Y_{ilb}Y_{jka} +2 \afive{\phi  
F}{ABa}\afive{\phi F}{ABb}Y_{ikb}Y_{jla}  
\nonumber\\&+2 \afive{\phi  
\widetilde{F}}{ABa}\afive{\phi  
\widetilde{F}}{ABb}Y_{ilb}Y_{jka} +2 \afive{\phi  
\widetilde{F}}{ABa}\afive{\phi  
\widetilde{F}}{ABb}Y_{ikb}Y_{jla}  
\nonumber\\&-\frac{8}{3}  
g\afive{\phi  
F}{ABa}Y_{ija}\Big[\afive{\psi F}{A}  
t^{B}\Big]_{kl} -\frac{8}{3} g\afive{\phi  
F}{ABa}Y_{ija}\Big[\afive{\psi F}{A}  
t^{B}\Big]_{lk}  
\nonumber\\&+\frac{8}{3}  
g\mathrm{i}\afive{\phi  
\widetilde{F}}{ABa}Y_{ija}\Big[\afive{\psi  
F}{A} t^{B}\Big]_{kl} +\frac{8}{3}  
g\mathrm{i}\afive{\phi  
\widetilde{F}}{ABa}Y_{ija}\Big[\afive{\psi  
F}{A} t^{B}\Big]_{lk} \nonumber\\&-4  
Y_{ija}\Big[\afive{\psi F}{A} Y^\dagger_{a}  
\afive{\psi F}{A}\Big]_{kl} -4  
Y_{ija}\Big[\afive{\psi F}{A} Y^\dagger_{a}  
\afive{\psi F}{A}\Big]_{lk}  
\nonumber\\&+24  
g^2\Big[\afive{\psi F}{A}  
t^{B}\Big]_{jk}\Big[\afive{\psi F}{B}  
t^{A}\Big]_{li} +24 g^2\Big[\afive{\psi F}{A}  
t^{B}\Big]_{ik}\Big[\afive{\psi F}{B}  
t^{A}\Big]_{lj} \nonumber\\&+48  
g^2\Big[\afive{\psi F}{A}  
t^{B}\Big]_{jk}\Big[\afive{\psi F}{B}  
t^{A}\Big]_{il} +48 g^2\Big[\afive{\psi F}{A}  
t^{B}\Big]_{ik}\Big[\afive{\psi F}{B}  
t^{A}\Big]_{jl} \nonumber\\&+24  
g^2\Big[\afive{\psi F}{A}  
t^{B}\Big]_{jl}\Big[\afive{\psi F}{B}  
t^{A}\Big]_{ki} +24 g^2\Big[\afive{\psi F}{A}  
t^{B}\Big]_{il}\Big[\afive{\psi F}{B}  
t^{A}\Big]_{kj} \nonumber\\&+32  
g^2\Big[\afive{\psi F}{A}  
t^{B}\Big]_{lk}\Big[(t^T)^{B} \afive{\psi  
F}{A}\Big]_{ij} -32 g^2\Big[\afive{\psi F}{A}  
t^{B}\Big]_{ij}\Big[\afive{\psi F}{A}  
t^{B}\Big]_{lk} \nonumber\\&-32  
g^2\Big[\afive{\psi F}{A}  
t^{B}\Big]_{ik}\Big[\afive{\psi F}{A}  
t^{B}\Big]_{jl} -32 g^2\Big[\afive{\psi F}{A}  
t^{B}\Big]_{il}\Big[\afive{\psi F}{A}  
t^{B}\Big]_{jk} \nonumber\\&+32  
g^2\Big[\afive{\psi F}{A}  
t^{B}\Big]_{kl}\Big[(t^T)^{B} \afive{\psi  
F}{A}\Big]_{ij} -32 g^2\Big[\afive{\psi F}{A}  
t^{B}\Big]_{ij}\Big[\afive{\psi F}{A}  
t^{B}\Big]_{kl}  
\nonumber\\&+\afive{\psi \phi^2  
}{ilab}\afive{\psi \phi^2 }{jkab}  
+\afive{\psi \phi^2 }{ikab}\afive{\psi  
\phi^2 }{jlab} \nonumber\\&+\frac{4}{3}  
\asix{\bar{\psi} \psi}{mknl}Y_{ija}Y_{mna}  
+\frac{5}{3} \asix{\bar{\psi}  
\psi}{mjnl}Y_{ima}Y_{kna}  
\nonumber\\&-\frac{2}{3}  
\asix{\bar{\psi} \psi}{njml}Y_{ima}Y_{kna}  
+\frac{5}{3} \asix{\bar{\psi}  
\psi}{minl}Y_{jma}Y_{kna}  
\nonumber\\&-\frac{2}{3}  
\asix{\bar{\psi} \psi}{niml}Y_{jma}Y_{kna}  
+\frac{5}{3} \asix{\bar{\psi}  
\psi}{mjnk}Y_{ima}Y_{lna}  
\nonumber\\&-\frac{2}{3}  
\asix{\bar{\psi} \psi}{njmk}Y_{ima}Y_{lna}  
+\frac{5}{3} \asix{\bar{\psi}  
\psi}{mink}Y_{jma}Y_{lna}  
\nonumber\\&-\frac{2}{3}  
\asix{\bar{\psi} \psi}{nimk}Y_{jma}Y_{lna} -4  
g\asix{\psi F}{Ajla}\Big[Y_{a}  
t^{A}\Big]_{ik} \nonumber\\&-4  
g\asix{\psi F}{Ajka}\Big[Y_{a}  
t^{A}\Big]_{il} +4 g\asix{\psi  
F}{Alja}\Big[Y_{a} t^{A}\Big]_{ik}  
\nonumber\\&+4 g\asix{\psi  
F}{Akja}\Big[Y_{a} t^{A}\Big]_{il} -4  
g\asix{\psi F}{Aila}\Big[(t^T)^{A}  
Y_{a}\Big]_{kj} \nonumber\\&-4  
g\asix{\psi F}{Aika}\Big[(t^T)^{A}  
Y_{a}\Big]_{lj} +4 g\asix{\psi  
F}{Alia}\Big[(t^T)^{A} Y_{a}\Big]_{kj}  
\nonumber\\&+4 g\asix{\psi  
F}{Akia}\Big[(t^T)^{A} Y_{a}\Big]_{lj}   
- g^2\asix{\psi  
\psi}{ijkm}\Big[t^{A} t^{A}\Big]_{ml}  
\nonumber\\&- g^2\asix{\psi  
\psi}{jikm}\Big[t^{A} t^{A}\Big]_{ml} -  
g^2\asix{\psi \psi}{ijlm}\Big[t^{A}  
t^{A}\Big]_{mk} \nonumber\\&-
g^2\asix{\psi \psi}{jilm}\Big[t^{A}  
t^{A}\Big]_{mk} -\frac{1}{3} \asix{\psi  
\psi}{mknl}Y_{ija}Y^\dagger_{mna}  
\nonumber\\&-\frac{1}{3}  
\asix{\psi \psi}{mlnk}Y_{ija}Y^\dagger_{mna}  
+\frac{1}{6} \asix{\psi  
\psi}{mjkl}\Big[Y_{a} Y^\dagger_{a}\Big]_{im}  
\nonumber\\&+\frac{1}{6}  
\asix{\psi \psi}{mikl}\Big[Y_{a}  
Y^\dagger_{a}\Big]_{jm} +\frac{1}{6}  
\asix{\psi \psi}{mjlk}\Big[Y_{a}  
Y^\dagger_{a}\Big]_{im}  
\nonumber\\&+\frac{1}{6}  
\asix{\psi \psi}{milk}\Big[Y_{a}  
Y^\dagger_{a}\Big]_{jm} -\frac{2}{3}  
g^2\asix{\psi \psi}{mjnl}t^{A}_{mi}t^{A}_{nk}  
\nonumber\\&-\frac{2}{3}  
g^2\asix{\psi \psi}{minl}t^{A}_{mj}t^{A}_{nk}  
+\frac{8}{3} g^2\asix{\psi  
\psi}{minj}t^{A}_{ml}t^{A}_{nk}  
\nonumber\\&+\frac{8}{3}  
g^2\asix{\psi \psi}{mjni}t^{A}_{ml}t^{A}_{nk}  
-\frac{2}{3} g^2\asix{\psi  
\psi}{mjnk}t^{A}_{mi}t^{A}_{nl}  
\nonumber\\&-\frac{2}{3}  
g^2\asix{\psi \psi}{mink}t^{A}_{mj}t^{A}_{nl}  
+\frac{8}{3} g^2\asix{\psi  
\psi}{minj}t^{A}_{mk}t^{A}_{nl}  
\nonumber\\&+\frac{8}{3}  
g^2\asix{\psi \psi}{mjni}t^{A}_{mk}t^{A}_{nl}
+(ij\leftrightarrow kl) 
-\frac{1}{2}(i\leftrightarrow k)\nonumber\\
&-\frac{1}{2}(i\leftrightarrow l)
-\frac{1}{2}(j\leftrightarrow k)
-\frac{1}{2}(j\leftrightarrow l)
.
\end{align}

\section{Conclusions \label{sec_concl}}
Effective field theories are widely adopted in particle physics, in particular in the context of physics beyond the Standard Model. Through the process of matching and running, the low-energy effects of heavy degrees of freedom are incorporated in a well-organized perturbative expansion.
This is particularly efficient for computing the implications of new heavy-physics models, especially since one can split the calculation into two independent parts, i.e. the bottom-up and top-down approaches.

Despite these processes being fully automated at one loop, the need to repeat the calculations for each different model compromises the efficiency of the EFT approach and makes it prone to errors. 
In a previous work \cite{Fonseca:2025zjb}, we presented an alternative solution based on computing the one-loop beta functions for a general effective Lagrangian, so that it encompasses any specific example and can be used after only a straightforward group-theoretical calculation. 
These beta functions were however calculated only for bosonic operators. 

In this follow-up work, we computed the beta functions for fermionic operators, completing the one-loop renormalization of the general effective Lagrangian up to dimension six.
Following the procedure introduced in \cite{Fonseca:2025zjb}, 
we derived the off-shell, one-loop divergences of the theory using the background field method, which we parametrize in a Green's basis of operators. Upon canonical normalization, we project the result onto a physical basis and provide the beta functions for all physical operators.

These new results for fermionic contributions have been cross-checked with  the leptonic sector of the SMEFT and the ALP EFT, as well as with several different toy models. 
In addition to being applicable to renormalize any EFT at one loop up to dimension six,
these results also set the ground for extending the general renormalization program, both in the direction of higher loop order or operator dimension.

\section*{Acknowledgments}

We would like to thank L. Bresciani, J.C. Criado, G. Guedes, J. F. Martin, A. E. Thomsen and N. Selimovic for useful discussions. This work has been partially supported by the Spanish Ministry of Science and Innovation
and SRA (10.13039/501100011033), with ERDF under grants PID2022-139466NB-C21 and PID2022-139466NB-C22,
by the Junta de Andaluc{\'\i}a grants FQM 101 and P21\_00199, by 
Consejer{\'\i}a de Universidad,
Investigaci{\'o}n e Innovaci{\'o}n,
Gobierno de Espa\~na 
and Uni{\'o}n Europea - NextGenerationEU
under grants AST22$\_$6.5 and CNS2024-154834.
This work has also received funding by the INFN Iniziative Specifiche APINE, from
the European Union's Horizon 2020 research and innovation programme under the Marie Sklodowska-Curie
grant agreement n. 101086085 - ASYMMETRY, and 
by the Italian MUR via the Departments of Excellence grant
2023- 2027 ``Quantum Frontiers'' and the PRIN 2022 project n. 2022K4B58X - AxionOrigins.

\appendix

\section{UV poles in the Green basis \label{matching:green}}

For completeness we report here the off-shell parametrization of the UV poles for fermionic operators after canonical normalization. 
\begin{align}
\ol
(m_\psi^\prime)_{ij} =&
\frac{3}{2}g^2 \Big[
m_\psi t^A t^A
\Big]_{ij}
-\frac{1}{4}\Big[m_\psi  Y^{\dagger}_a Y_a+2 Y_a m^\dagger_\psi Y_a\Big]_{ij}
+\frac{1}{4} \afive{\psi \phi^2}{ijab} (m^2_{\phi})_{ab}
\nonumber \\
&
+g \Big[6\afive{\psi F}{A} m_\psi^\dagger m_\psi t^A
+\frac{3}{2} m_\psi t^A m^\dagger_\psi \afive{\psi F}{A}
+\frac{3}{2} m_\psi \afive{\psi F}{A}^{\dagger} m_\psi t^A 
\Big]_{ij}
\nonumber \\ & 
+6 \Big[
m_\psi \afive{\psi F}{A}^{\dagger} m_\psi m^\dagger_\psi \afive{\psi F}{A}
+\afive{\psi F}{A} m^\dagger_\psi m_\psi m^\dagger_\psi \afive{\psi F}{A}
\Big]_{ij}
\nonumber \\ &
-\frac{1}{4}
\asix{\psi \psi}{ijkl} \Big[m^\dagger_\psi m_\psi m^\dagger_\psi\Big]_{kl}
-
\asix{\bar{\psi} \psi}{kilj} \Big[m_\psi m^\dagger_\psi m_\psi\Big]_{kl}
+(i\leftrightarrow j),
\\
\ol
(Y^\prime)_{ija}=&
-\frac{1}{8}(Y)_{ijb} \tr{Y_a Y^{\dagger}_b+Y_b Y^{\dagger}_a}
-\frac{1}{4} \Big[
Y_a Y_{b}^\dagger Y_b\Big]_{ij}
-\frac{1}{2} \Big[
Y_b Y^{\dagger}_a Y_b
\Big]_{ij}
+ \frac{3}{2} g^2 \Big[ Y_a t^A t^A\Big]_{ij}
\nonumber \\ &
-3 g^2 \Big[ t^{A\,T} m_\psi t^B\Big]_{ij} \afive{\phi F}{ABa}
+ \frac{3}{2} g \Big[
Y_a t^{A} m^\dagger_\psi \afive{\psi F}{A}
+Y_a \afive{\psi F}{A}^{\dagger} m_\psi t^A
\Big]_{ij}
\nonumber \\ &
+ 6 g \Big[
\afive{\psi F}{A} Y_a^{\dagger} m_\psi t^A
+\afive{\psi F}{A} m^\dagger_\psi  Y_a t^A
\Big]_{ij}
+\frac{1}{4} \afive{\psi \phi^2}{ijbc} \kappa_{abc}
+\afive{\psi \phi^2}{ikab} \Big[m^\dagger_{\psi} Y_b \Big]_{kj}
\nonumber \\ &
+12 g \Big[\afive{\psi F}{A} m^\dagger_\psi m_\psi t^B 
\Big]_{ij} \afive{\phi F}{ABa}
+ 6 g \asix{\psi F}{Aika} \Big[m^\dagger_\psi m_\psi t^A\Big]_{kj}
\nonumber \\ &
+6 \Big[Y_a \afive{\psi F}{A}^{\dagger} m_\psi m^\dagger_\psi \afive{\psi F}{A} \Big]_{ij}
+6 \Big[\afive{\psi F}{A} m^\dagger_\psi Y_a m^\dagger_\psi \afive{\psi F}{A}\Big]_{ij}
\nonumber \\ &
+12 \Big[
\afive{\psi F}{A} Y^\dagger_a m_\psi m^\dagger_\psi \afive{\psi F}{A}
\Big]_{ij} 
+\frac{1}{4} \asix{\psi \phi}{ijabc} (m_\phi^2)_{bc}
\nonumber \\ &
+\frac{1}{2} Y_{ijb} (m_{\phi}^2)_{cd} \asix{\phi D}{abcd}
-\mathrm{i} \asix{\phi \psi}{kiab} \Big\{\Big[
m_\psi m^\dagger_\psi Y_b\Big]_{kj} 
 + (m_\phi^2)_{bc} Y_{kjc} \Big\}
  \nonumber \\ &
- \asix{\bar{\psi}\psi}{kilj} \Big[m_\psi Y^\dagger_a m_\psi
  +2 Y_a m^\dagger_\psi m_\psi\Big]_{lk}
-\frac{1}{4} \asix{\psi \psi}{ijkl} \Big[m^\dagger_\psi Y_a m^\dagger_\psi
+2 Y^\dagger_a m_\psi m^\dagger_\psi \Big]_{lk}\nonumber\\&
+ (i \leftrightarrow j).
\end{align}

For dimension-five WCs:
\begin{align}
\ol
\afivep{\psi F}{Aij}=&
\Big[-\frac{1}{12} f^{ABC} f^{DBC} 
-\frac{1}{24} \tr{\theta^A\theta^D}
-\frac{1}{6}\tr{t^A t^D} \Big]g^2\afive{\psi F}{Dij}
\nonumber \\ &
- \frac{g^2}{2} \Big[\afive{\psi F}{B}t^A t^B\Big]_{ij}
-\frac{g^2}{2} \Big[\afive{\psi F}{A}t^Bt^B\Big]_{ij}
+\mathrm{i}\frac{7g^2}{2} \afive{\psi F}{B} f^{BAC}t^C\nonumber\\
&-\frac{1}{4}\Big[Y^a\overline{Y}^a \afive{\psi F}{A}\Big]_{ij}
\nonumber \\ &
-\frac{3g}{2} \Big[\afive{\psi F}{A}  \overline{\afive{\psi F}{B}} m_\psi t^B\Big]_{ij}
+\frac{g}{2} \tr{\overline{\afive{\psi F}{A}}m_\psi t^B
+\overline{\afive{\psi F}{B}}m_\psi t^A} \afive{\psi F}{Bij}
\nonumber \\
&
+\frac{3g}{2} \Big[\afive{\psi F}{A}  t^B m_\psi^\dagger \afive{\psi F}{B}\Big]_{ij}
-\frac{g}{2} \tr{t^{A}m_\psi^\dagger\afive{\psi F}{B} +t^{B}m_\psi^\dagger\afive{\psi F}{A} } \afive{\psi F}{Bij}
\nonumber\\
&
-\Big[\afive{\phi F}{ABa}+\mathrm{i}\afive{\phi \widetilde{F}}{ABa}\Big]
\Big[\afive{\psi F}{B} m_\psi^\dagger Y^a\Big]_{ij}
\nonumber\\
&-\mathrm{i}\frac{g^2}{4} 
\Big[
\asix{3F}{ABC}+\mathrm{i}\asix{3\widetilde{F}}{ABC}
\Big]
\Big[t^{B\,T} m_\psi t^C\Big]_{ij}
-\frac{g}{12}
\asix{\psi \psi}{ikjl}
\Big[m_\psi^\dagger t^{A\,T}\Big]_{kl}
\nonumber\\
&-\Big[Y^a m_\psi^\dagger \Big]_{ik} \asix{\psi F}{Akja} 
-(i\leftrightarrow j),
\\
\ol
\afivep{\psi \phi^2}{ijab} =&    \sum_{\mathrm{perm}}
\Big\{
\frac{1}{2}\Big[
g^2 \theta^A_{bc} \theta^A_{cd}-\frac{1}{4} \tr{Y^b \overline{Y}^d+Y^d \overline{Y}^b} 
\Big] \afive{\psi \phi^2}{ijad}
-\frac{1}{4} \Big[
g^2 t^A t^A
+\frac{1}{2} \overline{Y}^c Y^c
\Big]_{kj} \afive{\psi \phi^2}{ikab} 
\nonumber \\ &
\phantom{\sum_{\mathrm{perm}}\Big\{}
-\frac{g^2}{4}\theta^A_{ac} \theta^A_{bd} \afive{\psi \phi^2}{ijcd}
+g^2 \theta^A_{ac} t^A_{kj} \afive{\psi \phi^2}{ikbc}
-g^2 t^{A\,T}_{ik} \afive{\psi \phi^2}{klab} t^A_{lj}
\nonumber \\ &
\phantom{\sum_{\mathrm{perm}}\Big\{}
-\Big[Y^c \overline{Y}^a\Big]_{ik} \afive{\psi \phi^2}{kjbc}
-\frac{1}{8} \lambda_{abcd} \afive{\psi \phi^2}{ijcd}
-\frac{1}{4} Y^c_{ik} \afive{\psi \phi^2}{klab}^\dagger Y^c_{lj}
\nonumber \\ &
\phantom{\sum_{\mathrm{perm}}\Big\{}
-6 g \afive{\psi F}{Aik}\Big[g^2 \theta^A_{ac} \theta^B_{cb} t^B
+\overline{Y}^a Y^b t^A\Big]_{kj} 
+3g^2\Big[t^{A\,T}Y^a t^B\Big]_{ij} \afive{\phi F}{ABb}
\nonumber \\ &
\phantom{\sum_{\mathrm{perm}}\Big\{}
+6g^2\Big[t^{B\,T}m_\psi t^C\Big]_{ij} \afive{\phi F}{ABa}\afive{\phi F}{ACb}
-6g^2 \theta^A_{ac} \theta^B_{cb} \Big[\afive{\psi F}{A} \overline{m}_\psi \afive{\psi F}{B}\Big]_{ij}
\nonumber \\ &
\phantom{\sum_{\mathrm{perm}}\Big\{}
-\frac{3}{4} g 
\Big[
\overline{\afive{\psi F}{A}}m_\psi t^A
-t^A m^\dagger_\psi \afive{\psi F}{A}
\Big]_{kj}
 \afive{\psi \phi^2}{ikab}
\nonumber\\&
\phantom{\sum_{\mathrm{perm}}\Big\{}
+\frac{1}{2}\afive{\psi \phi^2}{ikac} \afive{\psi \phi^2}{jlbc} (m^\dagger_\psi)_{kl} 
\nonumber \\ &
\phantom{\sum_{\mathrm{perm}}\Big\{}
+3g \afive{\psi F}{Aik}\Big[
m^\dagger_\psi \afive{\psi\phi^2}{ab}t^A
+\afive{\psi \phi^2}{ab}^\dagger m_\psi t^A
\Big]_{kj}
\nonumber \\ &
\phantom{\sum_{\mathrm{perm}}\Big\{}
+12 g \Big[t^{B\,T}\Big(
m_\psi Y^\dagger_a+ Y_a m^\dagger_\psi
\Big)\Big]_{ik}\afive{\psi F}{Akj}\afive{\phi F}{ABb} 
\nonumber \\ &
\phantom{\sum_{\mathrm{perm}}\Big\{}
-6 \afive{\psi F}{Aik} \Big[
Y^\dagger_a m_\psi Y^\dagger_b+2 Y^\dagger_a Y_b m^\dagger_\psi
\Big]_{kl} \afive{\psi F}{Alj}
\nonumber \\ &
\phantom{\sum_{\mathrm{perm}}\Big\{}
+\frac{3}{2}g^2 \Big[
t^{A\,T} m_\psi t^B 
\Big]_{ij} \asix{\phi F}{ABab}
-\asix{\phi D}{abcd} \Big[
Y_c m^\dagger_\psi Y_d
\Big]_{ij}
\nonumber \\ &
\phantom{\sum_{\mathrm{perm}}\Big\{}
+\mathrm{i}\Big(
Y_{ikd} \kappa_{bcd}
+3 g^2 \theta^A_{bc}\Big[
 t^{A\,T}m_\psi \Big]_{ik} + 
\Big[Y_c Y^\dagger_b m_\psi
+Y_c m^\dagger_\psi Y_b\Big]_{ik}
\Big)
\asix{\phi \psi}{kjac}
\nonumber \\ &
\phantom{\sum_{\mathrm{perm}}\Big\{}
+\asix{\bar{\psi}\psi}{kilj} \Big[
 Y_a m^\dagger_\psi Y_b
+2Y_a Y^\dagger_b m_\psi\Big]_{kl}
-6g \asix{\psi F}{Aika} \Big[
Y^\dagger_b m_\psi t^A
+m^\dagger_\psi Y^b t^A
\Big]_{kj}
\nonumber \\ &
\phantom{\sum_{\mathrm{perm}}\Big\{}
-\frac{1}{4} \kappa_{acd}\asix{\psi \phi}{ijbcd}
-\frac{1}{2} \Big[Y_c m^\dagger_\psi \Big]_{ik}
\asix{\psi \phi}{kjabc}
\nonumber\\&
\phantom{\sum_{\mathrm{perm}}\Big\{}
+\frac{1}{4} \Big[
Y^\dagger_a m_\psi Y^\dagger_b
+2 Y^\dagger_a Y_b m^\dagger_\psi
\Big]_{kl}\asix{\psi \psi}{ijkl}
\Big\},
\\
\ol
\rfivep{\psi}{ij} =&\,
3g\Big[\afive{\psi F}{A}t^A\Big]_{ij}+(i\leftrightarrow j),
\\
\ol
\rfivep{\psi \phi}{ija} =&\,
3\mathrm{i} g^2 \Big[t^A t^B\Big]_{ij} \afive{\phi \widetilde{F}}{ABa}
-\frac{1}{2} \afive{\psi \phi^2}{ikab}^\dagger 
Y_{kjb}
\nonumber \\ &
+ 3 g \Big[
\overline{\afive{\psi F}{A}}t^{A\,T}Y_a
+\overline{\afive{\psi F}{A}}Y_a t^{A}
-\theta^A_{ab}Y^\dagger_b \afive{\psi F}{A}
- t^A Y^\dagger_a \afive{\psi F}{A}
\Big]_{ij}
\nonumber \\ &
-6g \Big[
t^A m^\dagger_\psi \afive{\psi F}{B}
\Big]_{ij} \afive{\phi F}{ABa}
+6\mathrm{i} g \Big[
t^A m^\dagger_\psi \afive{\psi F}{B}
-\overline{\afive{\psi F}{A}} m_\psi t^B
\Big]_{ij} \afive{\phi \widetilde{F}}{ABa}
\nonumber \\ &
+6 \Big[
\overline{\afive{\psi F}{A}} Y^a m^\dagger_\psi \afive{\psi F}{A}
+ \overline{\afive{\psi F}{A}} m_\psi Y^\dagger_a \afive{\psi F}{A}
\Big]_{ij}
\nonumber \\ &
+ 3 g \Big[\overline{\asix{\psi F}{Aa}}m_\psi t^A\Big]_{ij}
+ \mathrm{i} \Big[
\frac{3}{2}\asix{\phi \psi}{ab}
m^\dagger_\psi Y_b
+ Y^\dagger_b m_\psi \asix{\phi \psi}{ab}
\Big]_{ij}
\nonumber \\ &
+\asix{\bar{\psi}\psi}{ijkl} \Big[Y_a m^\dagger_\psi \Big]_{kl}
-\asix{\bar{\psi}\psi}{ijkl} \Big[m_\psi Y^\dagger_a \Big]_{kl}.
\end{align}
And, finally, the ones corresponding to dimension 6 operators,
\begin{align}
\ol
\asixp{\phi \psi}{ijab} =&
3g^2 [t^C t^B+t^B t^C]_{ij} \afive{\phi F}{ABa} \afive{\phi \widetilde{F}}{ACb}
\nonumber \\&
-3 \mathrm{i} g \theta^{A}_{ac} \afive{\phi F}{ABb} 
\Big[\afive{\psi F}{B}^\dagger Y_{c}
+Y_{c}^\dagger\afive{\psi F}{B} \Big]_{ij}
\nonumber \\ &
-3g \afive{\phi \widetilde{F}}{ABa}\Big[
t^A Y_b^\dagger \afive{\psi F}{B}
+\afive{\psi F}{B}^\dagger Y_b t^A
\Big]_{ij}
\nonumber \\ &
-\frac{3g}{2} \mathrm{i}
\theta^A_{bc}
\Big[ 
\afive{\psi F}{A}^\dagger  \afive{\psi \phi^2}{ac}
+\afive{\psi \phi^2}{ac}^\dagger  \afive{\psi F}{A}
\Big]_{ij}
\nonumber \\ &
+\frac{1}{8}\mathrm{i} \Big[
\afive{\psi \phi^2}{ac}^\dagger \afive{\psi \phi^2}{bc}
-\afive{\psi \phi^2}{bc}^\dagger \afive{\psi \phi^2}{ac}
\Big]_{ij}
\nonumber \\ &
+ 2 g^2 \Big[ 
\asix{\phi\psi}{ac} t^A - t^A \asix{\phi\psi}{ac}
\Big]_{ij} \theta^A_{bc}
\nonumber \\ &
-\frac{g^2}{4}  \Big[
\asix{\phi\psi}{ab}t^At^A+t^At^A\asix{\phi\psi}{ab}
\Big]_{ij}
+\frac{g^2}{2} \Big[t^A \asix{\phi \psi}{ab} t^A\Big]_{ij}
\nonumber \\ &
+ \Big[
\asix{\phi \psi}{bc}Y^\dagger_a Y_c + Y^\dagger_c Y_a \asix{\phi \psi}{bc}
\Big]_{ij}
-\frac{1}{8} \Big[ 
\asix{\phi\psi}{ab}Y^\dagger_{c} Y_c + Y^\dagger_c Y_c \asix{\phi\psi}{ab}
\Big]_{ij} 
\nonumber \\ &
-\frac{1}{4} \asix{\phi\psi}{ijac} \tr{Y_c Y^\dagger_b + Y_b Y^\dagger_c }
-\frac{1}{4}\Big[Y^\dagger_c \asix{\phi\psi}{ab}^T Y_c\Big]_{ij}
\nonumber \\ &
-\frac{3g}{2}\mathrm{i} \theta^A_{ac} 
\Big[
\asix{\psi F}{Ab}^\dagger Y_c
+ Y_c^\dagger \asix{\psi F}{Ab}
\Big]_{ij}
+\frac{3g}{2}\mathrm{i}
\Big[
t^A Y_b^\dagger \asix{\psi F}{Aa}
-\asix{\psi F}{Aa}^\dagger Y_b t^A
\Big]_{ij}
\nonumber \\ &
+2\mathrm{i} \asix{\phi D}{acbd}\Big[
Y_d^\dagger Y_c
\Big]_{ij}
+\mathrm{i} \asix{\bar{\psi}\psi}{ijkl} \Big[ Y_a Y^\dagger_b\Big]_{kl}
-(a \leftrightarrow b)
,\\
\ol
\asixp{\bar{\psi}\psi}{ijkl} =&
-6 g^2 
\Big[\afive{\psi F}{A}^\dagger \afive{\psi F}{B}\Big]_{ij}
\Big[3t^A t^B+t^B t^A\Big]_{kl}
-6 
\Big[ Y^\dagger_a \afive{\psi F}{A}\Big]_{ij}
\Big[ \afive{\psi F}{A}^\dagger Y_a \Big]_{kl}
\nonumber \\ &
-3 
\Big[\afive{\psi F}{A}^\dagger Y_a\Big]_{ij}
\Big[\afive{\psi F}{A}^\dagger Y_a\Big]_{kl}
-3 
\Big[Y^\dagger_a \afive{\psi F}{A}\Big]_{ij}
\Big[Y^\dagger_a \afive{\psi F}{A}\Big]_{kl}
\nonumber \\ &  
-\frac{g^2}{2} \asix{\bar{\psi}\psi}{ijkm} \Big[t^A t^A\Big]_{ml}
-\frac{g^2}{2} \asix{\bar{\psi}\psi}{ijml} \Big[t^A t^A\Big]_{km}
\nonumber \\ &
+g^2 \asix{\bar{\psi}\psi}{ijmn} t^A_{km} t^A_{nl}
+g^2 \asix{\bar{\psi}\psi}{ilmn} t^A_{km} t^A_{nj}
\nonumber \\ &
-2g^2 \asix{\bar{\psi}\psi}{imkn} t^A_{ml} t^A_{nj}
-2g^2 \asix{\bar{\psi}\psi}{mjnl} t^A_{km} t^A_{in}
\nonumber \\ &
-\frac{1}{2} Y^\dagger_{ina} Y_{jma} \asix{\bar{\psi}\psi}{mnkl}
\nonumber \\ &
-\frac{1}{4} \Big[Y^\dagger_{a} Y_{a}\Big]_{im} \asix{\bar{\psi}\psi}{mjkl}
-\frac{1}{4} \Big[Y^\dagger_{a} Y_{a}\Big]_{mj} \asix{\bar{\psi}\psi}{imkl}
\nonumber \\ &
+\frac{1}{12} Y_{jna} Y_{lma} \overline{\asix{\psi\psi}{mkni}}
+\frac{1}{12} Y^\dagger_{ina} Y^\dagger_{kma} \asix{\psi\psi}{mlnj}
\nonumber \\ &
-\mathrm{i} Y^\dagger_{ima} Y_{jmb} \asix{\phi \psi}{klab}
+(i \leftrightarrow k)(j \leftrightarrow l),\\
\ol
\asixp{\psi F}{Aija} =&
\afive{\phi F}{ABb} \Big\{
g^2\theta^C_{ab} \Big[ \afive{\psi F}{B} t^C \Big]_{ij}
+\frac{g^2}{2} \mathrm{i} f^{BCD}\theta^D_{ab} \afive{\psi F}{Cij}
- \Big[Y_b Y^\dagger_a \afive{\psi F}{B}\Big]_{ij}
\Big\}
\nonumber \\ &
+g^2 f^{BDE} f^{CDE} \afive{\phi F}{ABa} \afive{\psi F}{Cij} 
+\frac{5g^2}{2} f^{ACE} f^{BDE} \afive{\phi F}{BCa} \afive{\psi F}{Dij}
\nonumber \\ &
+3\mathrm{i} g^2 f^{BCD} \afive{\phi F}{ABa} 
\Big[\afive{\psi F}{C} t^D\Big]_{ij}
\nonumber \\ &
-9\mathrm{i} g^2 f^{ACD} \afive{\phi F}{BCa} 
\Big[\afive{\psi F}{D} t^B\Big]_{ij}
-g^2 \afive{\phi F}{BCa} 
\Big[ 
 \afive{\psi F}{B} t^A t^C
\Big]_{ij}
\nonumber \\ &
+\mathrm{i}\afive{\phi \widetilde{F}}{ABb} \Big\{
g^2\theta^C_{ab} \Big[ \afive{\psi F}{B} t^C \Big]_{ij}
+i\frac{g^2}{2}  f^{BCD}\theta^D_{ab} \afive{\psi F}{Cij}
- \Big[Y_b Y^\dagger_a \afive{\psi F}{B}\Big]_{ij}
\Big\} 
\nonumber \\ &
+\mathrm{i}g^2 f^{BDE} f^{CDE} \afive{\phi \widetilde{F}}{ABa} \afive{\psi F}{Cij} 
+\mathrm{i} g^2 f^{ACE} f^{BDE} \afive{\phi \widetilde{F}}{BCa} \afive{\psi F}{Dij}
\nonumber \\ &
-3 g^2 f^{BCD} \afive{\phi \widetilde{F}}{ABa} 
\Big[\afive{\psi F}{C} t^D\Big]_{ij}
\nonumber \\ &
+ 2 g^2 f^{ACD} \afive{\phi \widetilde{F}}{BCa} \afive{\psi F}{B} t^D
- \mathrm{i} g^2 \afive{\phi \widetilde{F}}{BCa}  \afive{\psi F}{B} t^A t^C
\nonumber \\ &
-2 \mathrm{i} g f^{ABC} \Big[ \afive{\psi F}{B}  Y^\dagger_a \afive{\psi F}{C}\Big]_{ij} 
-3 g \Big[
\afive{\psi F}{B} t^A Y^\dagger_a \afive{\psi F}{B}
\Big]_{ij}
\nonumber \\ &
+\frac{g}{2} \Big[ \afive{\phi F}{ABb} + \mathrm{i} \afive{\phi \widetilde{F}}{ABb}\Big] 
\Big[ \afive{\psi \phi^2}{ab} t^B\Big]_{ij}
- g \afive{\phi F}{BCa} \Big[ \afive{\phi F}{ABb} + \mathrm{i} \afive{\phi \widetilde{F}}{ABb} \Big]  \Big[Y_b t^C\Big]_{ij}
\nonumber \\ &
-\frac{1}{4} \Big[ Y_b Y^\dagger_b \asix{\psi F}{Aa}\Big]_{ij} 
- \Big[Y_b Y^\dagger_a \asix{\psi F}{Ab}\Big]_{ij}
-\frac{1}{8} \asix{\psi F}{Aijb} \tr{Y_b Y^\dagger_a+Y_a Y^\dagger_b}
\nonumber \\ &
+\frac{g^2}{2} \Big[ \theta^A_{ac} \theta^B_{bc} + \theta^B_{ac} \theta^A_{bc}\Big] \asix{\psi F}{Bijb}
-\frac{g^2}{2} \theta^B_{ac} \theta^B_{cb} \asix{\psi F}{Aijb}
+\frac{g^2}{24} \theta^A_{bc} \theta^B_{bc} \asix{\psi F}{Bija}
\nonumber \\ &
-\frac{g^2}{12} f^{ACD} f^{BCD} \asix{Bija}{}
+g^2 \theta^B_{ab} \Big[ \asix{\psi F}{Ab} t^B \Big]_{ij}
\nonumber \\ &
-\frac{7}{2} g^2 \mathrm{i} f^{ABC} \Big[\asix{\psi F}{Ba} t^C]_{ij}
- \frac{1}{2} g^2 
\Big[
\asix{\psi F}{Ba} t^A t^B
\Big]_{ij}
\nonumber\\&
- \frac{g^2}{2} 
\Big[
\asix{\psi F}{Aa} t^B t^B
\Big]_{ij}
- \frac{g^2}{6} 
\asix{\psi F}{Bija} \tr{t^A t^B}
\nonumber \\ &
-\frac{1}{2} \Big[\asix{\phi F}{ABab} + \mathrm{i} \asix{\phi \widetilde{F}}{ABab} \Big] 
\Big[ 
Y_b t^B
\Big]_{ij}
-\frac{g^2}{4}  \Big[\mathrm{i} \asix{3F}{ABC} - \asix{3\widetilde{F}}{ABC}\Big]
\Big[ t^{B\,T} Y_a t^C\Big]_{ij}
\nonumber \\ &  
-\frac{g}{12} \asix{\psi \psi}{kijl} \Big[t^A Y^\dagger_a\Big]_{lk}
-(i \leftrightarrow j),\\
\ol 
\asixp{\psi \phi}{ijabc} &=
\sum_{(ij),(abc)} \Bigg\{
12 g^3 \afive{\phi F}{ABa} \theta^B_{cd} \theta^C_{bd} \Big[ \afive{\psi F}{A} t^C +\afive{\psi F}{C} t^A\Big]_{ij}
\nonumber \\ &
\phantom{\sum_{(ij),(abc)} \Bigg\{}
-12 g \afive{\phi F}{ABc} \Big[ \afive{\psi F}{A} Y^\dagger_a Y_b t^B\Big]_{ij}
-\frac{3g^2}{2} \afive{\phi F}{ABc} \Big[ t^{A\, T} \afive{\psi \phi^2}{ab} t^B\Big]_{ij}
\nonumber \\ &
\phantom{\sum_{(ij),(abc)} \Bigg\{}
+3g \Big[
\afive{\psi F}{A} \afive{\psi \phi^2}{bc}^\dagger Y_a t^A
+\afive{\psi F}{A} Y^\dagger_a \afive{\psi \phi^2}{bc}  t^A
\Big]_{ij}
\nonumber \\ &
\phantom{\sum_{(ij),(abc)} \Bigg\{}
+\frac{1}{24} \afive{\phi}{abcde} \afive{\psi \phi^2}{ijde}
+\frac{1}{2} 
\Big[ 
\afive{\psi \phi^2}{cd} \afive{\psi \phi^2}{ab}^\dagger Y_d 
+ \afive{\psi \phi^2}{cd} Y_a^\dagger \afive{\psi \phi^2}{bd} 
\Big]_{ij}
\nonumber \\ &
\phantom{\sum_{(ij),(abc)} \Bigg\{}
+6 g^2 \afive{\phi F}{ABc} \afive{\phi F}{BCb} \Big[ t^{C\, T} Y_a t^A\Big]_{ij}
\nonumber \\ &
\phantom{\sum_{(ij),(abc)} \Bigg\{}
+6 g^2 \theta^A_{cd} \theta^B_{bd} \Big[ \afive{\psi F}{B} Y^\dagger_a \afive{\psi F}{A}\Big]_{ij}
-6 \Big[ \afive{\psi F}{A} Y^\dagger_a Y_b Y^\dagger_c \afive{\psi F}{A}\Big]_{ij}
\nonumber \\ &
\phantom{\sum_{(ij),(abc)} \Bigg\{}
+6 g^3 \theta^A_{cd} \theta^B_{bd} \Big[\asix{\psi F}{Aa} t^B\Big]_{ij}
-6 g \Big[
\asix{\psi F}{Aa} Y^\dagger_b Y_c t^A
\Big]_{ij}
\nonumber \\ &
\phantom{\sum_{(ij),(abc)} \Bigg\{}
-\frac{g^2}{4} \theta^A_{dc} \theta^A_{de} \asix{\psi \phi}{ijabe}
-\frac{g^2}{4} \theta^A_{bd} \theta^A_{ce} \asix{\psi \phi}{ijade}
-\frac{g^2}{2} \theta^A_{cd} \Big[\asix{\psi \phi}{abd} t^A\Big]_{ij}
\nonumber \\ &
\phantom{\sum_{(ij),(abc)} \Bigg\{}
+\frac{g^2}{4}\asix{\psi \phi}{abc} t^A t^A
-\frac{1}{8} \asix{\psi \phi}{ijade} \lambda_{bcde}
\nonumber \\ &
\phantom{\sum_{(ij),(abc)} \Bigg\{}
-\frac{1}{24} 
\Big[
2 Y_d \asix{\psi \phi}{abc}^\dagger Y_d
+ Y_d Y^\dagger_{d} \asix{\psi \phi}{abc}
+12 Y_d Y^\dagger_a \asix{\psi \phi}{bcd}
\Big]_{ij}
\nonumber \\ &
\phantom{\sum_{(ij),(abc)} \Bigg\{}
-\frac{1}{16} 
\asix{\psi \phi}{ijabd} \tr{Y^\dagger_c Y_d +Y^\dagger_d Y_c }
\nonumber \\ &
\phantom{\sum_{(ij),(abc)} \Bigg\{}
+3\mathrm{i} \lambda_{abde} \Big[ Y_e \asix{\phi \psi}{cd}\Big]_{ij}
+3\mathrm{i}g^2 \theta^A_{cd} \Big[ t^{A\, T} Y_a \asix{\phi \psi}{bd} \Big]_{ij}
+\mathrm{i} \Big[ Y_d Y^\dagger_a Y_b \asix{\phi \psi}{cd}\Big]_{ij}
\nonumber \\ &
\phantom{\sum_{(ij),(abc)} \Bigg\{}
+\frac{3g^2}{2} \asix{\phi F}{ABbc} \ferbil{t^{A\,T} Y_a t^B}{ij}
+2 \asix{\phi D}{bdce} \ferbil{Y_d Y^\dagger_a Y_e}{ij}
\nonumber \\ &
\phantom{\sum_{(ij),(abc)} \Bigg\{}
+ \asix{\bar{\psi}\psi}{kilj} \ferbil{Y_b Y^\dagger_c Y_a}{kl}
-\frac{1}{2} \asix{\psi \psi}{kilj} \ferbil{Y^\dagger_a Y_b Y^\dagger_c}{kl}
\Bigg\},\\
\ol
\asixp{\psi\psi}{ijkl}=& \Big[18 g^2 \Big[\afive{\psi F}{A} t^B\Big]_{ij}\Big[\afive{\psi F}{A} t^B\Big]_{kl}
    +18 g^2 \Big[\afive{\psi F}{A} t^B\Big]_{ij}\Big[\afive{\psi F}{B} t^A\Big]_{kl}\nonumber\\
    &+\frac{3}{16}\asix{\psi\phi^2}{ijab}\asix{\psi\phi^2}{klab}\nonumber\\
    &+\frac{3}{4} g^2 \asix{\psi\psi}{ijkm}\Big[t^A t^A\Big]_{ml} + \frac{3}{2}g^2\asix{\psi\psi}{mink} t^A_{mj} t^A_{nl} \nonumber\\
    & - \frac{1}{8} \asix{\psi\psi}{mjkl}\Big[Y^\dagger_a Y_a\Big]_{mi}\nonumber\\
    & -\frac{1}{4} \asix{\overline{\psi}\psi}{mknl} Y_{ima}Y_{jna}+\asix{\overline{\psi}\psi}{mjnl} Y_{ina} Y_{kma} \nonumber\\ 
    & +(i\leftrightarrow j) + (k\leftrightarrow l) + (i\leftrightarrow j)(k\leftrightarrow l)\Big]\Big[1 +(ij\leftrightarrow kl)\Big]\\
\ol
\rsixp{DF\psi}{Aij} =&
-3g \Big[
\afive{\psi F}{B}^\dagger \afive{\psi F}{A} t^B
\Big]_{ij}
-3g \Big[
t^B\afive{\psi F}{A}^\dagger \afive{\psi F}{B}
\Big]_{ij}
-\frac{2g}{3} \Big[
\afive{\psi F}{B} t^{A}\afive{\psi F}{B}^\dagger
\Big]_{ji} 
\nonumber \\ &
+\frac{17g}{3} \mathrm{i}
f^{ABC} \Big[\afive{\psi F}{B} \afive{\psi F}{C}^\dagger \Big]_{ji}
\nonumber \\ &
-2 \afive{\phi F}{ABa} \Big[
Y^\dagger_a \afive{\psi F}{B}+\afive{\psi F}{B}^\dagger Y_a
\Big]_{ij}
+\mathrm{i} \afive{\phi \widetilde{F}}{ABa}
\Big[
Y^\dagger_a \afive{\psi F}{B}
-\afive{\psi F}{B}^\dagger Y_a
\Big]_{ij}
\nonumber \\ &
+\mathrm{i}g^2 \asix{3F}{ABC} \Big[t^B t^C\Big]_{ij}
-\frac{g}{6} \mathrm{i} \theta^A_{ab} \asix{\phi \psi}{ijab}
-\frac{2g}{3} t^A_{lk} \asix{\bar{\psi}\psi}{ijkl}
-\frac{1}{2} \Big[
Y_a^\dagger \asix{\psi F}{Aa} + \asix{\psi F}{Aa}^\dagger Y_a
\Big]_{ij}
,\\
\ol
\rsixp{F\psi}{Aij} =&
\mathrm{i}g \Big[
 t^B \afive{\psi F}{A}^\dagger \afive{\psi F}{B}
-\afive{\psi F}{B}^\dagger \afive{\psi F}{A} t^B
\nonumber \\ &
-\afive{\phi \widetilde{F}}{ABa} \Big[ 
Y^\dagger_a \afive{\psi F}{B}
+\afive{\psi F}{B}^\dagger Y_a
\Big]_{ij}
+\frac{1}{2} \mathrm{i} \Big[ 
Y^\dagger_a \asix{\psi F}{Aa}
- \asix{\psi F}{Aa}^\dagger Y_a
\Big]_{ij}
,\\
\ol
\rsixp{\widetilde{F}\psi}{Aij} =&
g \Big[-
\afive{\psi F}{B}^\dagger \afive{\psi F}{A} t^B
-2 \afive{\psi F}{B}^\dagger t^{A\, T} \afive{\psi F}{B}
- t^B \afive{\psi F}{A}^\dagger \afive{\psi F}{B}
\Big]_{ij}
\nonumber \\ &
+4 g \mathrm{i} f^{ABC} 
\Big[
\afive{\psi F}{B}^\dagger \afive{\psi F}{C}
\Big]_{ij}
\nonumber \\ &
+\afive{\phi F}{ABa} 
\Big[
\afive{\psi F}{B}^\dagger Y_{a}
+Y^\dagger_{a} \afive{\psi F}{B}
\Big]_{ij}
+\frac{1}{2} \Big[
\asix{\psi F}{Aa}^\dagger Y_a
+Y^\dagger_a \asix{\psi F}{Aa}
\Big]_{ij}
,\\
\ol 
\rsixp{\phi \psi 1}{ijab} =&
3g^2 \theta^A_{ac} \theta^B_{bc} \Big[\afive{\psi F}{A}^\dagger \afive{\psi F}{B}\Big]_{ij}
-6 \Big[
\afive{\psi F}{A}^\dagger Y_a Y^\dagger_b \afive{\psi F}{A}
\Big]_{ij}
\nonumber \\ &
+\frac{3g}{4} \Big[ 
\afive{\psi F}{A}^\dagger \afive{\psi \phi^2}{ab} t^A
+ t^A \afive{\psi \phi^2}{ab}^\dagger \afive{\psi F}{A}
\Big]_{ij}
\nonumber \\ &
-3g \afive{\phi F}{ABa} \Big[ 
\afive{\psi F}{B}^\dagger Y_b t^A
+t^B Y^\dagger_b \afive{\psi F}{A}
\Big]_{ij}
+\frac{1}{4} \Big[
\afive{\psi \phi^2}{ac}^\dagger \afive{\psi \phi^2}{bc}
\Big]_{ij}
\nonumber \\ &
-\frac{1}{8} 
\Big[
\asix{\psi \phi }{abc}^\dagger Y_c
+Y^\dagger_c \asix{\psi \phi}{abc}
\Big]_{ij}
-\frac{3g}{2}
\Big[
\asix{\psi F}{Aa}^\dagger Y_b t^A
+ t^A Y^\dagger_b \asix{\psi F}{Aa}
\Big]_{ij}
\nonumber \\ &
+\frac{1}{4} \mathrm{i} 
\Big[
\asix{\phi \psi}{bc} Y^\dagger_a Y_c
-Y^\dagger_c Y_a \asix{\phi \psi}{bc}
\Big]_{ij}
+\asix{\phi D}{acbd} 
\Big[
Y^\dagger_d Y_c
\Big]_{ij}
+( a \leftrightarrow b),\\
\ol
\rsixp{\phi \psi 2}{ijab} &=
-\frac{3g}{2} \mathrm{i} \theta^A_{ac}
\Big[ 
\afive{\psi F}{A}^\dagger \afive{\psi \phi^2}{bc}
+\afive{\psi \phi^2}{bc}^\dagger \afive{\psi F}{A}
\Big]_{ij} 
\nonumber \\ &
+\frac{3g}{4} \mathrm{i} 
\Big[ 
\afive{\psi F}{A}^\dagger \afive{\psi \phi^2}{ab} t^A
- t^A \afive{\psi \phi^2}{ab}^\dagger \afive{\psi F}{A}
\Big]_{ij}
\nonumber \\ &
+3 g \mathrm{i} \theta^B_{bc} \afive{\phi F}{ABa} 
\Big[
\afive{\psi F}{A}^\dagger Y_c
+ Y^\dagger_c \afive{\psi F}{A}
\Big]_{ij}
\nonumber \\ &
+ 3g \mathrm{i} \afive{\phi F}{ABa}
\Big[
 t^B Y^\dagger_b \afive{\psi F}{A}
-\afive{\psi F}{A}^\dagger Y_b t^B
\Big]_{ij}
\nonumber \\ &
+3 g \afive{\phi \widetilde{F}}{ABa}
\Big[
t^A Y^\dagger_b \afive{\psi F}{B}
+ \afive{\psi F}{A}^\dagger Y_b t^B
\Big]_{ij}
\nonumber \\ &
+3g^2 \afive{\phi F}{ABa} \afive{\phi \widetilde{F}}{ACb}
\Big[
t^C t^B + t^B t^C
\Big]_{ij}
\nonumber \\ &
+\frac{3 g^2}{4} \theta^A_{bd} \theta^A_{cd} \asix{\phi \psi}{ijac}
+\frac{3 g^2}{4} \theta^A_{bc} 
\Big[
\asix{\phi \psi}{ac} t^A
- t^A \asix{\phi \psi}{ac}
\Big]_{ij}
\nonumber \\ &
+\frac{1}{4} 
\Big[
Y^\dagger_c Y_a \asix{\phi \psi}{bc}
+\asix{\phi \psi}{bc} Y^\dagger_a Y_c
\Big]_{ij}
+\frac{3g}{2} \mathrm{i}\theta^A_{bc} 
\Big[ 
\asix{\psi F}{Aa}^\dagger Y_c
+ Y^\dagger_c\asix{\psi F}{Aa} 
\Big]_{ij}
\nonumber \\ &
+\frac{3g^2}{2} \asix{\phi \widetilde{F}}{ABab} \Big[t^A t^B\Big]_{ij}  
+\frac{\mathrm{i}}{8} 
\Big[ 
\asix{\psi \phi}{abc}^\dagger Y_c 
- Y^\dagger_c \asix{\psi \phi}{abc}
\Big]_{ij}    
+( a \leftrightarrow b),\\
\ol 
\rsixp{\psi D}{ij} =&
-2 \Big[\afive{\psi F}{A}^\dagger \afive{\psi F}{A}\Big]_{ij}
,\\
\ol
\rsixp{\psi \phi D 1}{ija} =&
3 g \Big[\mathrm{i} \afive{\phi \widetilde{F}}{ABa} -\afive{\phi F}{ABa}   \Big]
\Big[
\afive{\psi F}{A} t^B  
\Big]_{ij}
-3 \Big[ \afive{\psi F}{A} Y^\dagger_a \afive{\psi F}{A}\Big]_{ij}
\nonumber \\ &
+\frac{1}{2}
\asix{\bar{\psi}\psi}{kilj} Y_{kla}
+\frac{\mathrm{i}}{4}
\Big[ Y_b \asix{\phi \psi}{ab}\Big]_{ij}
-\frac{1}{4} \asix{\psi \psi}{kilj} Y^\dagger_{kla}
+(i\leftrightarrow j),\\
\ol
\rsixp{\psi \phi D 2}{ija} =&
2g \afive{\phi F}{ABa} \Big[ \afive{\psi F}{A} t^B\Big]_{ij}
-2g \mathrm{i}  \afive{\phi \widetilde{F}}{ABa} \Big[ \afive{\psi F}{A} t^B\Big]_{ij}
\nonumber\\&
-\frac{\mathrm{i}}{2} \Big[ Y_b \asix{\phi \psi}{ab}\Big]_{ij} + (i \leftrightarrow j),\\
\ol
\rsixp{\psi \phi D 3}{ija} =&
-6 g \afive{\phi F}{ABa} \Big[ \afive{\psi F}{A} t^B - t^{B\,T} \afive{\psi F}{A}\Big]_{ij}
\nonumber \\ &
+ 6 g \Big[t^{A\,T} \asix{\psi F}{Aa}\Big]_{ij}
+\frac{3}{2} \mathrm{i} \Big[ Y_b \asix{\phi \psi}{ab} \Big]_{ij}
-\frac{1}{2} \mathrm{i} \Big[ Y_b \asix{\phi \psi}{ab} \Big]_{ji}.
\end{align}
The divergences for the evanescent operators read:
\begin{align}
    \ol \rsixp{\psi\overline{\psi}}{ijkl}=& 12 g^2 \Big[\afive{\psi F}{A} t^B\Big]_{ij}\Big[ t^A\afive{\psi F}{B}^\dagger\Big]_{kl}
    +12 g^2 \Big[\afive{\psi F}{A} t^B\Big]_{ij}\Big[ t^B\afive{\psi F}{A}^\dagger\Big]_{kl}\nonumber\\
    &+\frac{1}{8} \afive{\psi\phi^2}{ijab}\afive{\psi\phi^2}{klab}^\dagger\nonumber\\
    &\big[-\frac{1}{12} Y^\dagger_{kma}Y^\dagger_{lna} \asix{\psi\psi}{imjn} + \mathrm{h.c.}\big] \nonumber\\
    &- \frac{1}{2} \asix{\overline{\psi}\psi}{mjkn} Y_{ima} Y^\dagger_{nla} \nonumber\\
    &+(i\leftrightarrow j) + (k \leftrightarrow l)+(i \leftrightarrow j)(k \leftrightarrow l),\\
    \ol \rsixp{\psi\overline{\psi} 2}{ijkl}=&- \frac{1}{8} \asix{\overline{\psi}\psi}{mjkn} Y_{ima} Y^\dagger_{nla} \nonumber\\
    &+\frac{g}{4} \Big[ Y_a t^A\Big]_{ij} \asix{\psi F}{Akla}^\ast
    -\frac{g}{4} \asix{\psi F}{Aija} \Big[ t^A Y^\dagger\Big]_{kl} \nonumber \\
    &-(i\leftrightarrow j) - (k \leftrightarrow l)+(i \leftrightarrow j)(k \leftrightarrow l),\\
     \ol \rsixp{\psi\psi 2}{ijkl}=&-\frac{3}{2} g^2 \Big[\afive{\psi F}{A} t^B\Big]_{ij}\Big[\afive{\psi F}{A} t^B\Big]_{kl} \nonumber\\
     &+\frac{g^2}{4}\asix{\psi\psi}{mink}t^A_{mj}t^A_{nl}
     -\frac{3}{16} \asix{\overline{\psi}\psi}{njml}Y_{kna} Y_{ima}
     \nonumber\\
     &-\frac{3g}{4} \Big[Y_{a} t^A\Big]_{ij} \asix{\psi F}{Akla} \nonumber\\
     &+(i \leftrightarrow k)(j \leftrightarrow l)\nonumber\\
     &+\Big[1 + (i \leftrightarrow k)(j \leftrightarrow l)\Big]\Big[-(i\leftrightarrow j) - (k \leftrightarrow l)+(i \leftrightarrow j)(k \leftrightarrow l)\Big].
\end{align}

\bibliographystyle{style}
\bibliography{references}

@article{Machacek:1983tz,
    author = "Machacek, Marie E. and Vaughn, Michael T.",
    title = "{Two loop renormalization group equations in a general quantum field theory. 1. Wave function renormalization}",
    reportNumber = "NUB-2590, HUTP-83/A003",
    doi = "10.1016/0550-3213(83)90610-7",
    journal = "Nucl. Phys. B",
    volume = "222",
    pages = "83--103",
    year = "1983"
}

@article{Machacek:1983fi,
    author = "Machacek, Marie E. and Vaughn, Michael T.",
    title = "{Two loop renormalization group equations in a general quantum field theory. 2. Yukawa couplings}",
    reportNumber = "NUB 2611",
    doi = "10.1016/0550-3213(84)90533-9",
    journal = "Nucl. Phys. B",
    volume = "236",
    pages = "221--232",
    year = "1984"
}

@article{Jenkins:2013zja,
    author = "Jenkins, Elizabeth E. and Manohar, Aneesh V. and Trott, Michael",
    title = "{Renormalization Group Evolution of the Standard Model Dimension Six Operators I: Formalism and lambda Dependence}",
    eprint = "1308.2627",
    archivePrefix = "arXiv",
    primaryClass = "hep-ph",
    doi = "10.1007/JHEP10(2013)087",
    journal = "JHEP",
    volume = "10",
    pages = "087",
    year = "2013"
}

@article{Jenkins:2013wua,
    author = "Jenkins, Elizabeth E. and Manohar, Aneesh V. and Trott, Michael",
    title = "{Renormalization Group Evolution of the Standard Model Dimension Six Operators II: Yukawa Dependence}",
    eprint = "1310.4838",
    archivePrefix = "arXiv",
    primaryClass = "hep-ph",
    reportNumber = "CERN-PH-TH/2015-247",
    doi = "10.1007/JHEP01(2014)035",
    journal = "JHEP",
    volume = "01",
    pages = "035",
    year = "2014"
}

@article{Alonso:2013hga,
    author = "Alonso, Rodrigo and Jenkins, Elizabeth E. and Manohar, Aneesh V. and Trott, Michael",
    title = "{Renormalization Group Evolution of the Standard Model Dimension Six Operators III: Gauge Coupling Dependence and Phenomenology}",
    eprint = "1312.2014",
    archivePrefix = "arXiv",
    primaryClass = "hep-ph",
    reportNumber = "CERN-PH-TH-2013-305, CERN-PH-TH/2013-305",
    doi = "10.1007/JHEP04(2014)159",
    journal = "JHEP",
    volume = "04",
    pages = "159",
    year = "2014"
}

@article{Jenkins:2017dyc,
    author = "Jenkins, Elizabeth E. and Manohar, Aneesh V. and Stoffer, Peter",
    title = "{Low-Energy Effective Field Theory below the Electroweak Scale: Anomalous Dimensions}",
    eprint = "1711.05270",
    archivePrefix = "arXiv",
    primaryClass = "hep-ph",
    doi = "10.1007/JHEP01(2018)084",
    journal = "JHEP",
    volume = "01",
    pages = "084",
    year = "2018",
    note = "[Erratum: JHEP 12, 042 (2023)]"
}

@article{Chala:2021pll,
    author = "Chala, Mikael and Guedes, Guilherme and Ramos, Maria and Santiago, Jose",
    title = "{Towards the renormalisation of the Standard Model effective field theory to dimension eight: Bosonic interactions I}",
    eprint = "2106.05291",
    archivePrefix = "arXiv",
    primaryClass = "hep-ph",
    doi = "10.21468/SciPostPhys.11.3.065",
    journal = "SciPost Phys.",
    volume = "11",
    pages = "065",
    year = "2021"
}

@article{DasBakshi:2022mwk,
    author = "Das Bakshi, Supratim and Chala, Mikael and D\'\i{}az-Carmona, \'Alvaro and Guedes, Guilherme",
    title = "{Towards the renormalisation of the Standard Model effective field theory to dimension eight: bosonic interactions II}",
    eprint = "2205.03301",
    archivePrefix = "arXiv",
    primaryClass = "hep-ph",
    doi = "10.1140/epjp/s13360-022-03194-5",
    journal = "Eur. Phys. J. Plus",
    volume = "137",
    number = "8",
    pages = "973",
    year = "2022"
}

@article{Bakshi:2024wzz,
    author = "Bakshi, S. D. and Chala, M. and D\'\i{}az-Carmona, \'A. and Ren, Z. and Vilches, F.",
    title = "{Renormalization of the SMEFT to dimension eight: Fermionic interactions I}",
    eprint = "2409.15408",
    archivePrefix = "arXiv",
    primaryClass = "hep-ph",
    doi = "10.1007/JHEP12(2024)214",
    journal = "JHEP",
    volume = "12",
    pages = "214",
    year = "2025"
}

@article{AccettulliHuber:2021uoa,
    author = "Accettulli Huber, Manuel and De Angelis, Stefano",
    title = "{Standard Model EFTs via on-shell methods}",
    eprint = "2108.03669",
    archivePrefix = "arXiv",
    primaryClass = "hep-th",
    reportNumber = "QMUL-PH-21-32, SAGEX-21-17-E",
    doi = "10.1007/JHEP11(2021)221",
    journal = "JHEP",
    volume = "11",
    pages = "221",
    year = "2021"
}

@article{Assi:2023zid,
    author = "Assi, Beno\^\i{}t and Helset, Andreas and Manohar, Aneesh V. and Pag\`es, Julie and Shen, Chia-Hsien",
    title = "{Fermion geometry and the renormalization of the Standard Model Effective Field Theory}",
    eprint = "2307.03187",
    archivePrefix = "arXiv",
    primaryClass = "hep-ph",
    reportNumber = "CALT-TH-2023-024, FERMILAB-PUB-23-362-T",
    doi = "10.1007/JHEP11(2023)201",
    journal = "JHEP",
    volume = "11",
    pages = "201",
    year = "2023"
}

@article{Helset:2022pde,
    author = "Helset, Andreas and Jenkins, Elizabeth E. and Manohar, Aneesh V.",
    title = "{Renormalization of the Standard Model Effective Field Theory from geometry}",
    eprint = "2212.03253",
    archivePrefix = "arXiv",
    primaryClass = "hep-ph",
    reportNumber = "CALT-TH-2022-041",
    doi = "10.1007/JHEP02(2023)063",
    journal = "JHEP",
    volume = "02",
    pages = "063",
    year = "2023"
}

@article{Boughezal:2024zqa,
    author = "Boughezal, Radja and Huang, Yingsheng and Petriello, Frank",
    title = "{Renormalization-group running of dimension-8 four-fermion operators in the SMEFT}",
    eprint = "2408.15378",
    archivePrefix = "arXiv",
    primaryClass = "hep-ph",
    doi = "10.1103/PhysRevD.110.116015",
    journal = "Phys. Rev. D",
    volume = "110",
    number = "11",
    pages = "116015",
    year = "2024"
}

@article{Chala:2020wvs,
    author = "Chala, Mikael and Guedes, Guilherme and Ramos, Maria and Santiago, Jose",
    title = "{Running in the ALPs}",
    eprint = "2012.09017",
    archivePrefix = "arXiv",
    primaryClass = "hep-ph",
    doi = "10.1140/epjc/s10052-021-08968-2",
    journal = "Eur. Phys. J. C",
    volume = "81",
    number = "2",
    pages = "181",
    year = "2021"
}

@article{Bauer:2020jbp,
    author = "Bauer, Martin and Neubert, Matthias and Renner, Sophie and Schnubel, Marvin and Thamm, Andrea",
    title = "{The Low-Energy Effective Theory of Axions and ALPs}",
    eprint = "2012.12272",
    archivePrefix = "arXiv",
    primaryClass = "hep-ph",
    reportNumber = "IPPP/20/69, MITP/20-070 SISSA 30/2020/FISI, ZH-TH-47/20",
    doi = "10.1007/JHEP04(2021)063",
    journal = "JHEP",
    volume = "04",
    pages = "063",
    year = "2021"
}

@article{Bonilla:2021ufe,
    author = "Bonilla, J. and Brivio, I. and Gavela, M. B. and Sanz, V.",
    title = "{One-loop corrections to ALP couplings}",
    eprint = "2107.11392",
    archivePrefix = "arXiv",
    primaryClass = "hep-ph",
    reportNumber = "IFT-UAM/CSIC-21-82",
    doi = "10.1007/JHEP11(2021)168",
    journal = "JHEP",
    volume = "11",
    pages = "168",
    year = "2021"
}

@article{DiNoi:2024ajj,
    author = {Di Noi, Stefano and Gr\"ober, Ramona and Mandal, Manoj K.},
    title = "{Two-loop running effects in Higgs physics in Standard Model Effective Field Theory}",
    eprint = "2408.03252",
    archivePrefix = "arXiv",
    primaryClass = "hep-ph",
    reportNumber = "COMETA-2024-19",
    doi = "10.1007/JHEP12(2024)220",
    journal = "JHEP",
    volume = "12",
    pages = "220",
    year = "2025"
}

@article{Carmona:2021xtq,
    author = "Carmona, Adrian and Lazopoulos, Achilleas and Olgoso, Pablo and Santiago, Jose",
    title = "{Matchmakereft: automated tree-level and one-loop matching}",
    eprint = "2112.10787",
    archivePrefix = "arXiv",
    primaryClass = "hep-ph",
    doi = "10.21468/SciPostPhys.12.6.198",
    journal = "SciPost Phys.",
    volume = "12",
    number = "6",
    pages = "198",
    year = "2022"
}

@article{Fuentes-Martin:2022jrf,
    author = {Fuentes-Mart\'\i{}n, Javier and K\"onig, Matthias and Pag\`es, Julie and Thomsen, Anders Eller and Wilsch, Felix},
    title = "{A proof of concept for matchete: an automated tool for matching effective theories}",
    eprint = "2212.04510",
    archivePrefix = "arXiv",
    primaryClass = "hep-ph",
    reportNumber = "MITP-22-105, TUM-HEP-1443/22, ZU-TH-58/22",
    doi = "10.1140/epjc/s10052-023-11726-1",
    journal = "Eur. Phys. J. C",
    volume = "83",
    number = "7",
    pages = "662",
    year = "2023"
}

@article{Machacek:1984zw,
    author = "Machacek, Marie E. and Vaughn, Michael T.",
    title = "{Two loop renormalization group equations in a general quantum field theory. 3. Scalar quartic couplings}",
    reportNumber = "NUB-2653-REV, NUB-2653",
    doi = "10.1016/0550-3213(85)90040-9",
    journal = "Nucl. Phys. B",
    volume = "249",
    pages = "70--92",
    year = "1985"
}

@article{Luo:2002ti,
    author = "Luo, Ming-xing and Wang, Hua-wen and Xiao, Yong",
    title = "{Two loop renormalization group equations in general gauge field theories}",
    eprint = "hep-ph/0211440",
    archivePrefix = "arXiv",
    doi = "10.1103/PhysRevD.67.065019",
    journal = "Phys. Rev. D",
    volume = "67",
    pages = "065019",
    year = "2003"
}

@article{Fonseca:2025zjb,
    author = "Fonseca, Renato M. and Olgoso, Pablo and Santiago, Jos{\'e}",
    title = "{Renormalization of general Effective Field Theories: formalism and renormalization of bosonic operators}",
    eprint = "2501.13185",
    archivePrefix = "arXiv",
    primaryClass = "hep-ph",
    doi = "10.1007/JHEP07(2025)135",
    journal = "JHEP",
    volume = "07",
    pages = "135",
    year = "2025"
}

@article{Martin:1993zk,
    author = "Martin, Stephen P. and Vaughn, Michael T.",
    title = "{Two loop renormalization group equations for soft supersymmetry breaking couplings}",
    eprint = "hep-ph/9311340",
    archivePrefix = "arXiv",
    reportNumber = "NUB-3081-93-TH",
    doi = "10.1103/PhysRevD.50.2282",
    journal = "Phys. Rev. D",
    volume = "50",
    pages = "2282",
    year = "1994",
    note = "[Erratum: Phys.Rev.D 78, 039903 (2008)]"
}

@article{Fuentes-Martin:2020zaz,
    author = "Fuentes-Martin, Javier and Ruiz-Femenia, Pedro and Vicente, Avelino and Virto, Javier",
    title = "{DsixTools 2.0: The Effective Field Theory Toolkit}",
    eprint = "2010.16341",
    archivePrefix = "arXiv",
    primaryClass = "hep-ph",
    reportNumber = "MITP/20-061, IFIC/20-50",
    doi = "10.1140/epjc/s10052-020-08778-y",
    journal = "Eur. Phys. J. C",
    volume = "81",
    number = "2",
    pages = "167",
    year = "2021"
}

@article{DiNoi:2022ejg,
    author = "Di Noi, Stefano and Silvestrini, Luca",
    title = "{RGESolver: a C++ library to perform renormalization group evolution in the Standard Model Effective Theory}",
    eprint = "2210.06838",
    archivePrefix = "arXiv",
    primaryClass = "hep-ph",
    doi = "10.1140/epjc/s10052-023-11189-4",
    journal = "Eur. Phys. J. C",
    volume = "83",
    number = "3",
    pages = "200",
    year = "2023"
}

@article{Liao:2024xel,
    author = "Liao, Yi and Ma, Xiao-Dong and Wang, Hao-Lin",
    title = "{Probing dimension-8 SMEFT operators through neutral meson mixing}",
    eprint = "2409.10305",
    archivePrefix = "arXiv",
    primaryClass = "hep-ph",
    doi = "10.1007/JHEP03(2025)133",
    journal = "JHEP",
    volume = "03",
    pages = "133",
    year = "2025"
}

@article{Naterop:2024cfx,
    author = "Naterop, Luca and Stoffer, Peter",
    title = "{Renormalization-group equations of the LEFT at two loops: dimension-five effects}",
    eprint = "2412.13251",
    archivePrefix = "arXiv",
    primaryClass = "hep-ph",
    reportNumber = "PSI-PR-24-30, ZU-TH 66/24",
    doi = "10.1007/JHEP06(2025)007",
    journal = "JHEP",
    volume = "06",
    pages = "007",
    year = "2025"
}

@article{Naterop:2025lzc,
    author = "Naterop, Luca and Stoffer, Peter",
    title = "{Renormalization-group equations of the LEFT at two loops: dimension-six baryon-number-violating operators}",
    eprint = "2505.03871",
    archivePrefix = "arXiv",
    primaryClass = "hep-ph",
    reportNumber = "PSI-PR-25-09, ZU-TH 31/25, INT-PUB-25-013",
    doi = "10.1007/JHEP07(2025)237",
    journal = "JHEP",
    volume = "07",
    pages = "237",
    year = "2025"
}

@article{Naterop:2025cwg,
    author = "Naterop, Luca and Stoffer, Peter",
    title = "{Renormalization-group equations of the LEFT at two loops: dimension-six operators}",
    eprint = "2507.08926",
    archivePrefix = "arXiv",
    primaryClass = "hep-ph",
    reportNumber = "ZU-TH 48/25",
    month = "7",
    year = "2025"
}

@article{DiNoi:2025tka,
    author = {Di Noi, Stefano and Erdelyi, Barbara Anna and Gr{\"o}ber, Ramona},
    title = "{Complete two-loop Yukawa-induced running of the Higgs-gluon coupling in SMEFT}",
    eprint = "2510.14680",
    archivePrefix = "arXiv",
    primaryClass = "hep-ph",
    reportNumber = "KA-TP-31-2025, COMETA-2025-47",
    month = "10",
    year = "2025"
}

@article{Born:2024mgz,
    author = "Born, Lukas and Fuentes-Mart{\'\i}n, Javier and Kvedarait{\.{e}}, Sandra and Thomsen, Anders Eller",
    title = "{Two-loop running in the bosonic SMEFT using functional methods}",
    eprint = "2410.07320",
    archivePrefix = "arXiv",
    primaryClass = "hep-ph",
    doi = "10.1007/JHEP05(2025)121",
    journal = "JHEP",
    volume = "05",
    pages = "121",
    year = "2025"
}

@article{Haisch:2025lvd,
    author = "Haisch, Ulrich",
    title = "{Higgs production from anomalous gluon dynamics}",
    eprint = "2503.06249",
    archivePrefix = "arXiv",
    primaryClass = "hep-ph",
    reportNumber = "MPP-2025-38",
    doi = "10.1007/JHEP06(2025)004",
    journal = "JHEP",
    volume = "06",
    pages = "004",
    year = "2025"
}

@article{Haisch:2025vqj,
    author = "Haisch, Ulrich and Niggetiedt, Marco",
    title = "{Precision tests of third-generation four-quark operators: $gg \to h$ and $h \to \gamma \gamma$}",
    eprint = "2507.20803",
    archivePrefix = "arXiv",
    primaryClass = "hep-ph",
    reportNumber = "MPP-2025-147",
    month = "7",
    year = "2025"
}

@article{Guedes:2025sax,
    author = "Guedes, Guilherme and Roosmale Nepveu, Jasper",
    title = "{Two-loop renormalization of general bosonic effective field theories}",
    eprint = "2512.08827",
    archivePrefix = "arXiv",
    primaryClass = "hep-ph",
    reportNumber = "CERN-TH-2025-250",
    month = "12",
    year = "2025"
}

@article{delAguila:2000rc,
    author = "del Aguila, F. and Perez-Victoria, M. and Santiago, Jose",
    title = "{Observable contributions of new exotic quarks to quark mixing}",
    eprint = "hep-ph/0007316",
    archivePrefix = "arXiv",
    reportNumber = "UG-FT-118-00, MIT-CTP-2997",
    doi = "10.1088/1126-6708/2000/09/011",
    journal = "JHEP",
    volume = "09",
    pages = "011",
    year = "2000"
}

@article{delAguila:2008pw,
    author = "del Aguila, F. and de Blas, J. and Perez-Victoria, M.",
    title = "{Effects of new leptons in Electroweak Precision Data}",
    eprint = "0803.4008",
    archivePrefix = "arXiv",
    primaryClass = "hep-ph",
    reportNumber = "UG-FT-224-08, CAFPE-94-08",
    doi = "10.1103/PhysRevD.78.013010",
    journal = "Phys. Rev. D",
    volume = "78",
    pages = "013010",
    year = "2008"
}

@article{delAguila:2010mx,
    author = "del Aguila, F. and de Blas, J. and Perez-Victoria, M.",
    title = "{Electroweak Limits on General New Vector Bosons}",
    eprint = "1005.3998",
    archivePrefix = "arXiv",
    primaryClass = "hep-ph",
    reportNumber = "UG-FT-272-10, CAFPE-142-10, UG-FT-272/10, CAFPE-142/10",
    doi = "10.1007/JHEP09(2010)033",
    journal = "JHEP",
    volume = "09",
    pages = "033",
    year = "2010"
}

@article{deBlas:2014mba,
    author = "de Blas, Jorge and Chala, Mikael and Perez-Victoria, Manuel and Santiago, Jose",
    title = "{Observable Effects of General New Scalar Particles}",
    eprint = "1412.8480",
    archivePrefix = "arXiv",
    primaryClass = "hep-ph",
    reportNumber = "CERN-PH-TH-2014-264, DESY-15-011",
    doi = "10.1007/JHEP04(2015)078",
    journal = "JHEP",
    volume = "04",
    pages = "078",
    year = "2015"
}

@article{deBlas:2017xtg,
    author = "de Blas, J. and Criado, J. C. and Perez-Victoria, M. and Santiago, J.",
    title = "{Effective description of general extensions of the Standard Model: the complete tree-level dictionary}",
    eprint = "1711.10391",
    archivePrefix = "arXiv",
    primaryClass = "hep-ph",
    reportNumber = "CERN-TH-2017-251",
    doi = "10.1007/JHEP03(2018)109",
    journal = "JHEP",
    volume = "03",
    pages = "109",
    year = "2018"
}

@article{Guedes:2023azv,
    author = "Guedes, Guilherme and Olgoso, Pablo and Santiago, Jos{\'e}",
    title = "{Towards the one loop IR/UV dictionary in the SMEFT: One loop generated operators from new scalars and fermions}",
    eprint = "2303.16965",
    archivePrefix = "arXiv",
    primaryClass = "hep-ph",
    reportNumber = "DESY-23-040",
    doi = "10.21468/SciPostPhys.15.4.143",
    journal = "SciPost Phys.",
    volume = "15",
    number = "4",
    pages = "143",
    year = "2023"
}

@article{Guedes:2024vuf,
    author = "Guedes, Guilherme and Olgoso, Pablo",
    title = "{From the EFT to the UV: the complete SMEFT one-loop dictionary}",
    eprint = "2412.14253",
    archivePrefix = "arXiv",
    primaryClass = "hep-ph",
    reportNumber = "DESY-24-2",
    month = "12",
    year = "2024"
}

@article{Misiak:2025xzq,
    author = "Misiak, Miko{\l}aj and Na{\l}{\k{e}}cz, Ignacy",
    title = "{One-loop renormalization group equations in generic effective field theories. Part I. Bosonic operators}",
    eprint = "2501.17134",
    archivePrefix = "arXiv",
    primaryClass = "hep-ph",
    doi = "10.1007/JHEP06(2025)210",
    journal = "JHEP",
    volume = "06",
    pages = "210",
    year = "2025"
}

@article{Aebischer:2025zxg,
    author = "Aebischer, Jason and Bresciani, Luigi C. and Selimovic, Nudzeim",
    title = "{Anomalous dimension of a general effective gauge theory. Part I. Bosonic sector}",
    eprint = "2502.14030",
    archivePrefix = "arXiv",
    primaryClass = "hep-ph",
    reportNumber = "CERN-TH-2025-032",
    doi = "10.1007/JHEP08(2025)209",
    journal = "JHEP",
    volume = "08",
    pages = "209",
    year = "2025"
}

@article{Fonseca:2020vke,
    author = "Fonseca, Renato M.",
    title = "{GroupMath: A Mathematica package for group theory calculations}",
    eprint = "2011.01764",
    archivePrefix = "arXiv",
    primaryClass = "hep-th",
    doi = "10.1016/j.cpc.2021.108085",
    journal = "Comput. Phys. Commun.",
    volume = "267",
    pages = "108085",
    year = "2021"
}

@article{Abbott:1980hw,
    author = "Abbott, L. F.",
    title = "{The Background Field Method Beyond One Loop}",
    reportNumber = "CERN-TH-2973",
    doi = "10.1016/0550-3213(81)90371-0",
    journal = "Nucl. Phys. B",
    volume = "185",
    pages = "189--203",
    year = "1981"
}

@article{Chala:2025crd,
    author = "Chala, Mikael and L{\'o}pez Miras, Javier",
    title = "{New insights into two-loop running in effective field theories}",
    eprint = "2512.04064",
    archivePrefix = "arXiv",
    primaryClass = "hep-ph",
    month = "12",
    year = "2025"
}

\end{document}